\documentclass[aps,prb,twocolumn,floatfix,reprint]{revtex4-2}
\usepackage{amsmath,amssymb,bm}
\usepackage{graphicx}
\usepackage{booktabs}
\usepackage[dvipsnames]{xcolor}
\usepackage[colorlinks=true,linkcolor=blue!60!black,citecolor=green!40!black,urlcolor=blue!60!black]{hyperref}

\newcommand{\vk}{\mathbf{k}}
\newcommand{\vq}{\mathbf{q}}
\newcommand{\va}{\mathbf{a}}

\newcommand{\vM}{\mathbf{M}}
\newcommand{\vQ}{\mathbf{Q}}

\begin{document}
\title{Precursor pseudogap from charge fluctuations in kagome metals}
\author{Yuan Tian}
\author{Sergey Y. Savrasov}
\affiliation{Department of Physics, University of California, Davis, California 95616, USA}
\date{\today}

\begin{abstract}
Slow charge fluctuations near a density-wave instability reshape electronic states, but whether they open a precursor pseudogap or merely broaden the spectrum depends on the geometry of the Fermi surface and on the statistics of the fluctuations. We address this question for the kagome lattice at van Hove filling, where the Fermi surface is a hexagon with parallel faces whose sliding nesting is known to control the instabilities of the metal. Together with the sublattice structure of the Bloch states, it gives the charge susceptibility near the instability one of two forms: a \emph{ridge} along a continuous line in momentum space, or a \emph{peak} at isolated points. Describing both by an overdamped boson model, we obtain closed-form one-loop self-energies: the peak gaps the van Hove saddle points, the ridge the face midpoints. Determinant quantum Monte Carlo confirms both forms, and in the random-phase approximation, nearest-neighbor exchange produces a peak-form, bond-charge instability. We apply the theory to study charge fluctuations in CsV$_3$Sb$_5$, where many experiments suggest they are quasi-static. We treat them exactly with a sign-problem-free Monte Carlo simulation of an effective electron--lattice model. It reproduces the observed bond order and, just above the transition, a partial pseudogap at the van Hove saddle points. More generally, we conclude that whenever the ordering wave vectors close into a triangle, the local gap at the connected saddle points is set by the product of the three order-parameter components, so that a precursor pseudogap requires all three to coexist spatially.
\end{abstract}

\maketitle

%=====================================================================
\section{Introduction}
%=====================================================================

The kagome lattice combines, in its simplest tight-binding description, a flat band, Dirac cones and van Hove saddle points of the dispersion at the M points of the Brillouin zone. When the Fermi level lies at one of these saddle points the density of states diverges logarithmically, and the Fermi surface becomes a perfect hexagon whose corners are the saddle points. This van Hove filling has long been recognized as a fertile setting for electronic instabilities. Early work proposed topological phases~\cite{Guo2009} and unconventional superconductivity in doped kagome systems~\cite{Ko2009}, and chiral superconducting and chiral spin-density-wave states in the kagome Hubbard model~\cite{Yu2012}. Kiesel and Thomale pointed out that the Bloch states at the saddle points of the upper van Hove filling are localized on a single sublattice each, so that local interactions cannot efficiently scatter electrons between different saddle points; this \emph{sublattice interference} reshapes the competition between orders~\cite{Kiesel2012}. Functional renormalization group studies mapped a rich phase diagram of the extended Hubbard model at van Hove filling, including ferromagnetism, intra-unit-cell antiferromagnetism, charge and spin bond order, Pomeranchuk instabilities, and $s$-, $d$+$id$- and $f$-wave superconductivity~\cite{Kiesel2013,Wang2013}, and the kagome lattice was proposed as a host of strongly correlated Dirac metals~\cite{Mazin2014}.

Interest intensified with the discovery of the vanadium antimonides AV$_3$Sb$_5$ (A = K, Rb, Cs)~\cite{Ortiz2019,Ortiz2020,Ortiz2021K,Yin2021}, reviewed in Refs.~\cite{Neupert2022,Jiang2022,Wilson2024}. These layered metals contain nearly ideal vanadium kagome planes with van Hove singularities close to the Fermi level~\cite{Kang2022,Hu2022}, undergo a charge-density-wave (CDW) transition near 80--100\,K into a $2\times2$ in-plane (and $2\times2\times2$ or $2\times2\times4$ three-dimensional) superstructure~\cite{Ortiz2021X,Li2021PRX}, and become superconducting at a few kelvin, with a double superconducting dome under pressure~\cite{Chen2021,Yu2021,Zheng2022}. Much about the charge order remains debated. Its time-reversal properties are contested: muon-spin relaxation reports time-reversal-symmetry breaking at the CDW transition~\cite{Mielke2022,Khasanov2022}, whereas high-resolution Kerr measurements find no spontaneous signal~\cite{Saykin2023}. Rotational-symmetry breaking and nematicity have been reported~\cite{Li2022NP,Nie2022}. The transition is weakly first order, and the role of the lattice is itself debated. Early inelastic x-ray scattering found no phonon anomaly~\cite{Li2021PRX}, and diffuse scattering found quasi-static, order--disorder-type precursor fluctuations dominated by a central peak~\cite{Subires2023}. Recent high-resolution inelastic x-ray scattering combined with anharmonic first-principles calculations instead identifies a soft phonon along M--L, strongest at the L point, as the driver of the transition~\cite{McGuinness2026}, with the large electron--phonon linewidth of the soft mode offered as the reason it escaped earlier detection~\cite{Alkorta2023}. On the theory side, mean-field and Landau analyses of the van Hove saddle points~\cite{Denner2021,Park2021,Lin2021,Feng2021}, first-principles phonon calculations~\cite{Tan2021}, renormalization-group studies~\cite{Wu2021,Profe2024}, the random-phase approximation~\cite{Wu2024} and vertex corrections beyond it~\cite{Tazai2022} variously emphasize electronic and lattice mechanisms, real and imaginary bond orders, and the role of sublattice interference. Upon hole doping the long-range order disappears, but quasi-one-dimensional charge correlations~\cite{Kautzsch2023} and charge-order fluctuations with picosecond correlation times~\cite{Kongruengkit2025} persist well beyond the CDW phase boundary~\cite{Oey2022}.

These observations raise the question of how slow charge fluctuations near a density-wave instability scatter electrons, which they do most strongly where the Fermi surface is nested and the density of states is large, here at the van Hove saddle points, and what decides whether this scattering merely broadens the electronic states or opens a precursor pseudogap. Near a continuous instability the dynamical interaction between electrons is dominated by a soft, Landau-damped collective mode. To a good approximation it takes a simple relaxational form, an \emph{overdamped boson} peaked at the ordering wave vector. Scattering off such a boson concentrates on the parts of the Fermi surface connected by the ordering vector (so-called hot spots) and, close to the instability, can produce precursor pseudogaps and anisotropic quasiparticle damping. On the kagome lattice there is a twist. Because the faces of the hexagonal Fermi surface are straight and parallel, the set of nesting vectors is not a set of points but a set of lines, and in the site-charge channel the polarizability forms \emph{ridges} along $\Gamma$--M rather than peaks at M, whereas in the bond-charge channel it peaks at M. Which form the boson takes, and what each form does to the electrons, depends on the channel and on the sublattice structure of the fluctuations.

Overdamped boson models of this kind have been studied for more than three decades. The spin response proposed by Millis, Monien and Pines~\cite{MMP1990} to account for nuclear relaxation in the cuprates became the pairing interaction of the nearly antiferromagnetic Fermi liquid, in which Monthoux, Balatsky and Pines found $d$-wave superconductivity~\cite{Monthoux1991,Monthoux1993}. It had been established earlier for the Hubbard model that antiferromagnetic fluctuations favor $d$-wave pairing~\cite{Scalapino1986} (for a review, see Ref.~\cite{Scalapino2012}). Because the boson is concentrated near its ordering vector, its scattering is strongest at hot spots, the Fermi-surface points connected by that vector~\cite{Hlubina1995}, and close to the instability its thermally excited part opens precursor pseudogaps there~\cite{Vilk1997,Schmalian1998}. The spin-fermion model built on these ideas was developed into a controlled theory of the normal state near the quantum critical point and of its pairing instability~\cite{Abanov2003,Abanov2001}, and renormalization-group work showed that quasiparticles at the hot spots lose their Fermi-liquid character~\cite{Metlitski2010}. For charge order, singular overdamped interactions near incommensurate density-wave instabilities were analyzed by Castellani, Di Castro and Grilli~\cite{Castellani1995,Perali1996}, an approach later applied to the charge-density fluctuations seen by resonant x-ray scattering in the cuprates~\cite{Caprara2017,Arpaia2019}. The relaxational frequency dependence coincides with that of Hertz--Millis theory~\cite{Hertz1976,Millis1993}, while exact treatments of static fluctuations date back to the fluctuating Peierls chain of Lee, Rice and Anderson~\cite{LRA1973} and to Sadovskii's solution~\cite{Sadovskii1979}. Most recently, a charge boson of this type, whose scattering is confined to lines of a three-dimensional Fermi surface, was proposed as the source of the electron--boson coupling that density-functional calculations miss in Ba$_{1-x}$K$_x$BiO$_3$~\cite{WanSavrasov}.

In this work we use the overdamped boson model to analyze the charge instability, which on the kagome lattice at van Hove filling involves three ordering wave vectors connecting the three van Hove saddle points pairwise. The order is therefore described by a three-component Ginzburg--Landau order parameter $(\Phi_1,\Phi_2,\Phi_3)$, and because the three vectors add up to a reciprocal lattice vector, $\vQ_1+\vQ_2+\vQ_3=\mathbf G$, its free energy contains the cubic invariant $\Phi_1\Phi_2\Phi_3$~\cite{Denner2021,Park2021,Christensen2021}. Our central result, which we demonstrate with Monte Carlo simulations in which the coupled problem of electrons and slow charge fluctuations is approximated by an effective electron--lattice model, is that whether the fluctuations of this order open a precursor pseudogap depends on their statistics, not only on their strength or correlation length. The local gap at the saddle points is set by the product $\Phi_1\Phi_2\Phi_3$ and vanishes wherever any one component is small, so that a pseudogap requires the three components to coexist spatially. Gaussian fluctuations, which share the two-point correlations of the order but leave the components independent, therefore only broaden the states, whereas fluctuations with locally ordered domains open a pseudogap. Although our discussion focuses on the AV$_3$Sb$_5$ kagome metals, the same features should hold for triangular and honeycomb lattices that exhibit similar three-component ($2\times2$) charge order, such as the transition-metal dichalcogenide 1$T$-TiSe$_2$~\cite{DiSalvo1976}. This contrasts with one-component order parameters, such as Peierls order in one dimension or single-$\vQ$ density waves, for which Gaussian fluctuations with a long enough correlation length already produce a pseudogap~\cite{LRA1973,Sadovskii1979,Bartosch1999,Vilk1997,Schmalian1998}.

The rest of the paper is organized as follows. In Sec.~\ref{sec:model} we introduce the tight-binding model, sliding nesting, and the non-interacting site and bond susceptibilities; formulate the two forms of the overdamped boson, ridge and peak; and derive closed-form one-loop self-energies for $\delta$-function interactions. In Sec.~\ref{sec:rpa} we obtain both forms from the random-phase approximation for the extended Hubbard model, extract their parameters, compute the resulting single-particle spectra, and test these results with determinant quantum Monte Carlo. In Sec.~\ref{sec:static} we collect the experimental evidence that the charge fluctuations of CsV$_3$Sb$_5$ are quasi-static, present the Monte Carlo simulations that demonstrate our central result (Sec.~\ref{sec:scmc}), and identify the origin of the pseudogap at the saddle points (Sec.~\ref{sec:criterion}). Section~\ref{sec:conclusions} summarizes our conclusions.

%=====================================================================
\section{Model, sliding nesting and the overdamped boson}
\label{sec:model}
%=====================================================================

\subsection{Tight-binding model, nesting, and non-interacting susceptibilities}
\label{sec:tb}

\begin{figure*}[t]
\centering
\includegraphics[width=\textwidth]{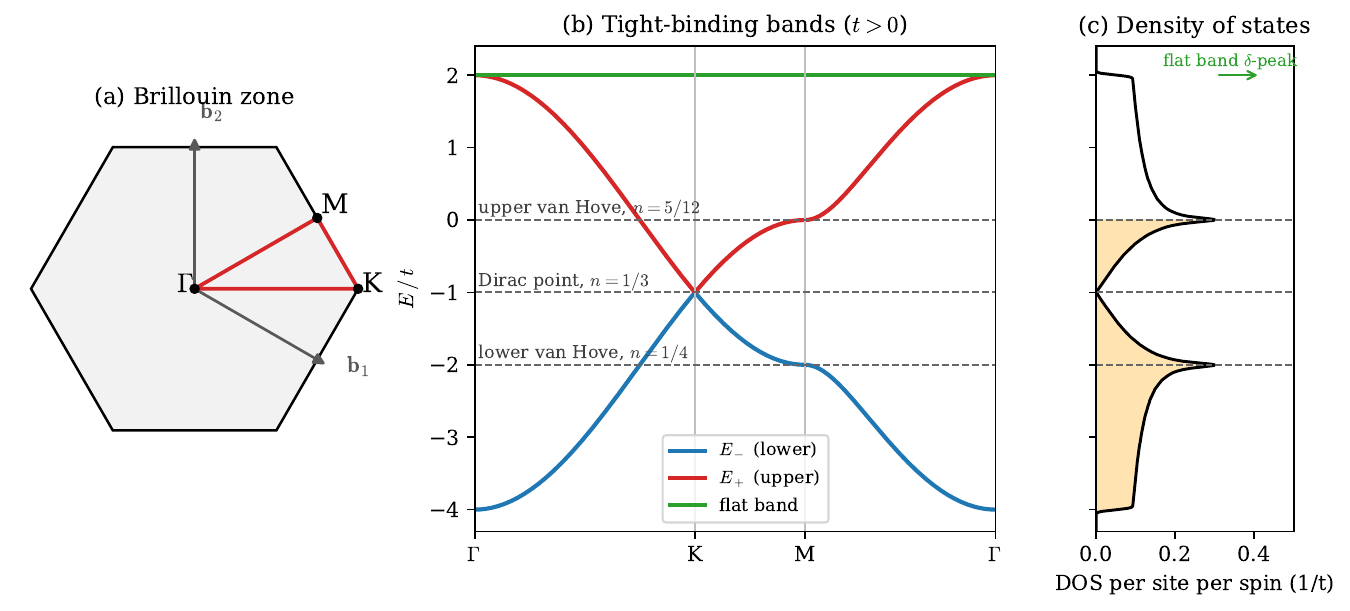}
\caption{(a) Brillouin zone and high-symmetry points. (b) Tight-binding bands on the kagome lattice; dashed lines mark the lower van Hove singularity ($-2t$), the Dirac point ($-t$) and the upper van Hove singularity ($0$). (c) Density of states of the dispersive bands; the shaded area is occupied at the upper van Hove filling.}
\label{fig:bands}
\end{figure*}

\paragraph{Bands.} We consider the kagome lattice with nearest-neighbor hopping $t$ (Appendix~\ref{app:tb}). Its nearest-neighbor bonds form three families of straight chains, A--B along $\va_1$, A--C along $\va_2$ and B--C along $\va_3$. Of the three bands, one is flat at $E=2t$; the two dispersive bands touch at the Dirac points K ($E=-t$) and have saddle points at the three M points, at $E=0$ (upper band) and $E=-2t$ (lower band) (Fig.~\ref{fig:bands}). The corresponding fillings are $n=5/12$ and $1/4$ per site and spin. For CsV$_3$Sb$_5$, a minimal fit to the vanadium bands found from density-functional calculations gives $t\approx0.365$\,eV~\cite{Barman2024}.

\begin{figure*}[t]
\centering
\includegraphics[width=\textwidth]{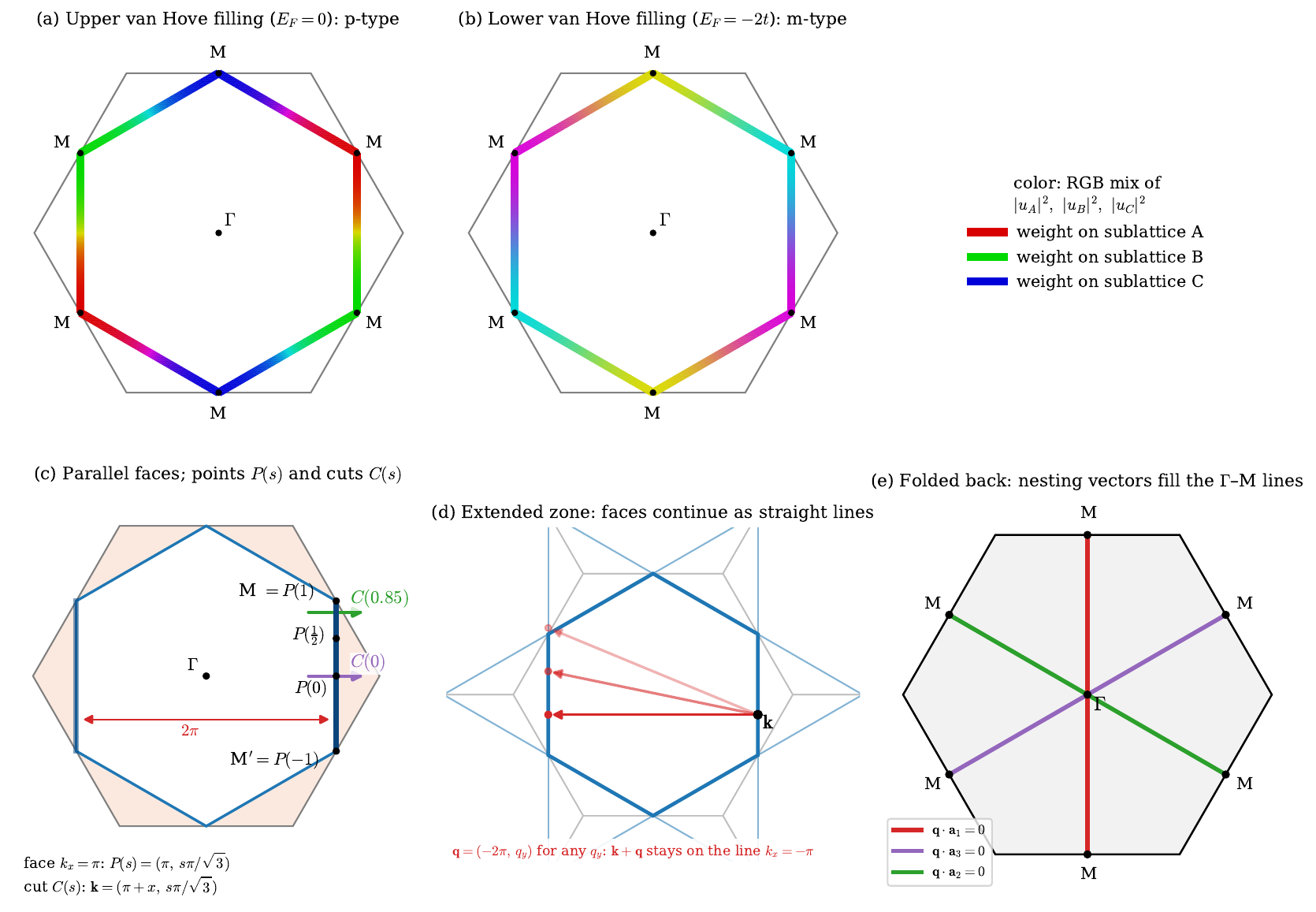}
\caption{Sublattice texture and sliding nesting. (a,b) Fermi surface colored by the sublattice weights of its Bloch states (red, green and blue for sublattices A, B and C): (a) upper van Hove filling, p-type (pure) saddle points; (b) lower van Hove filling, m-type (mixed) saddle points. (c) Opposite faces of the hexagon are parallel and separated by $2\pi$; on the face $k_x=\pi$, points $P(s)$ run from M$'=P(-1)$ through the face midpoint $P(0)$ to M $=P(1)$, and cuts $C(s)$ cross the face at $P(s)$. (d) In the extended zone the faces continue as straight lines, and a state $\vk$ on the face $k_x=\pi$ is carried onto the line $k_x=-\pi$ by $\vq=(-2\pi,q_y)$ for any $q_y$. (e) Folded back, the nesting vectors fill the three $\Gamma$--M lines $\vq\cdot\va_i=0$.}
\label{fig:nesting}
\end{figure*}

\paragraph{Fermi surface, sublattice texture and sliding nesting.} At either van Hove filling the Fermi surface is a perfect hexagon with corners at the M points; in the extended zone it consists of the complete straight lines $\vk\cdot\va_i=\pm\pi$ (Appendix~\ref{app:tb}). Its Bloch states have a characteristic sublattice texture~\cite{Kiesel2012} (Fig.~\ref{fig:nesting}a,b; Appendix~\ref{app:tb}): at the upper filling each saddle point is carried by a single sublattice, different for the three inequivalent M points, whereas at the lower filling the saddle states are shared equally between two sublattices. Following Refs.~\cite{Wu2021,Kang2022}, we call these van Hove singularities of \emph{pure} (p-type) and \emph{mixed} (m-type) sublattice character. At the p-type (upper) filling, a local vertex between two different saddle points vanishes because the two states reside on different sublattices; this is the sublattice interference of Ref.~\cite{Kiesel2012}.

Opposite faces of the hexagon are parallel and separated by $2\pi$ (Fig.~\ref{fig:nesting}c). The vector that carries a face onto the opposite one, $(2\pi,0)$ for the faces $k_x=\pm\pi$, folds back onto an M point. Because the faces are straight, adding any component parallel to the faces slides the shifted face along the opposite one without leaving the Fermi surface (Fig.~\ref{fig:nesting}d,e). The set of nesting vectors is therefore not the three M points but the three lines
\begin{equation}
\vq\cdot\va_i=0,\qquad i=1,2,3,
\label{eq:lines}
\end{equation}
which pass through $\Gamma$ and the M points. We refer to this as \emph{sliding nesting}. Near a face, $\xi_{\vk+\vq}\simeq-\xi_\vk$ for every $\vq$ on the corresponding line. The M points are distinguished only in that they also connect saddle point to saddle point. It is convenient to label points on the face $k_x=\pi$ as $P(s)=(\pi,\,s\pi/\sqrt3)$, with $s=0$ the face midpoint and $s=\pm1$ the M corners, and cuts perpendicular to it as $C(s):\ \vk=(\pi+x,\,s\pi/\sqrt3)$ (Fig.~\ref{fig:nesting}c). In this notation $\xi_\vk\simeq-x\cos(\pi s/2)$: the Fermi velocity perpendicular to the face, $v_\perp=\cos(\pi s/2)$ in units of $ta$, vanishes at the saddle points.

\paragraph{Non-interacting susceptibilities and vertices.} A charge-channel particle--hole operator with momentum $\vq$, $O_F(\vq)=\sum_\vk\sum_{ab}c^\dagger_{\vk+\vq,a}F_{ab}(\vk,\vq)c_{\vk b}$, is specified by a $3\times3$ sublattice form factor $F$: diagonal for the site charges $n_\alpha$, and off-diagonal, carrying the bond phases, for the bond-charge ($B^+$) and bond-current ($B^-$) operators on the six nearest-neighbor bonds of the unit cell. Their explicit forms, fifteen operators in total, are given in Appendix~\ref{app:tb}; a general fluctuation pattern is a normalized combination of them. In the band basis the operator acts through the vertex
\begin{equation}
\begin{split}
g^F_{\nu\mu}(\vk,\vk+\vq)&=\langle u^\nu(\vk+\vq)|F(\vk,\vq)|u^\mu(\vk)\rangle\\
&=\sum_{ab}u^{\nu}_a(\vk+\vq)\,F_{ab}(\vk,\vq)\,u^\mu_b(\vk),
\end{split}
\label{eq:vertex}
\end{equation}
in which the sublattice indices are contracted with the Bloch vectors $u^\mu(\vk)$ of the band $\mu$ (Appendix~\ref{app:tb}) while the band indices remain. This contraction is the origin of sublattice interference: for a site pattern, whose form factor is diagonal, $F=\mathrm{diag}(e_A,e_B,e_C)$, with entries $e_\alpha$ the amplitudes of the charge modulation on the sublattices $\alpha=A,B,C$, $g^F_{\nu\mu}=\sum_\alpha e_\alpha u^\nu_\alpha(\vk+\vq)u^\mu_\alpha(\vk)$ vanishes between Bloch states that reside on different sublattices. The bare polarizability of two such operators is (per spin)
\begin{equation}
\begin{split}
\chi^0_{FF'}(\vq,\omega)={}&\frac1N\sum_{\vk}\sum_{\mu\nu}\frac{f(E_\mu(\vk))-f(E_\nu(\vk+\vq))}{E_\nu(\vk+\vq)-E_\mu(\vk)-\omega-i0^+}\\
&\times g^F_{\nu\mu}(\vk,\vk+\vq)\,\big[g^{F'}_{\nu\mu}(\vk,\vk+\vq)\big]^*
\end{split}
\label{eq:chi0}
\end{equation}
For the three site densities this is the $3\times3$ site polarizability $\chi^0_{\alpha\beta}$; with all fifteen operators of Eq.~\eqref{eq:formfactors} it is a $15\times15$ matrix (Sec.~\ref{sec:rpaA}).

\begin{figure*}[t]
\centering
\includegraphics[width=\textwidth]{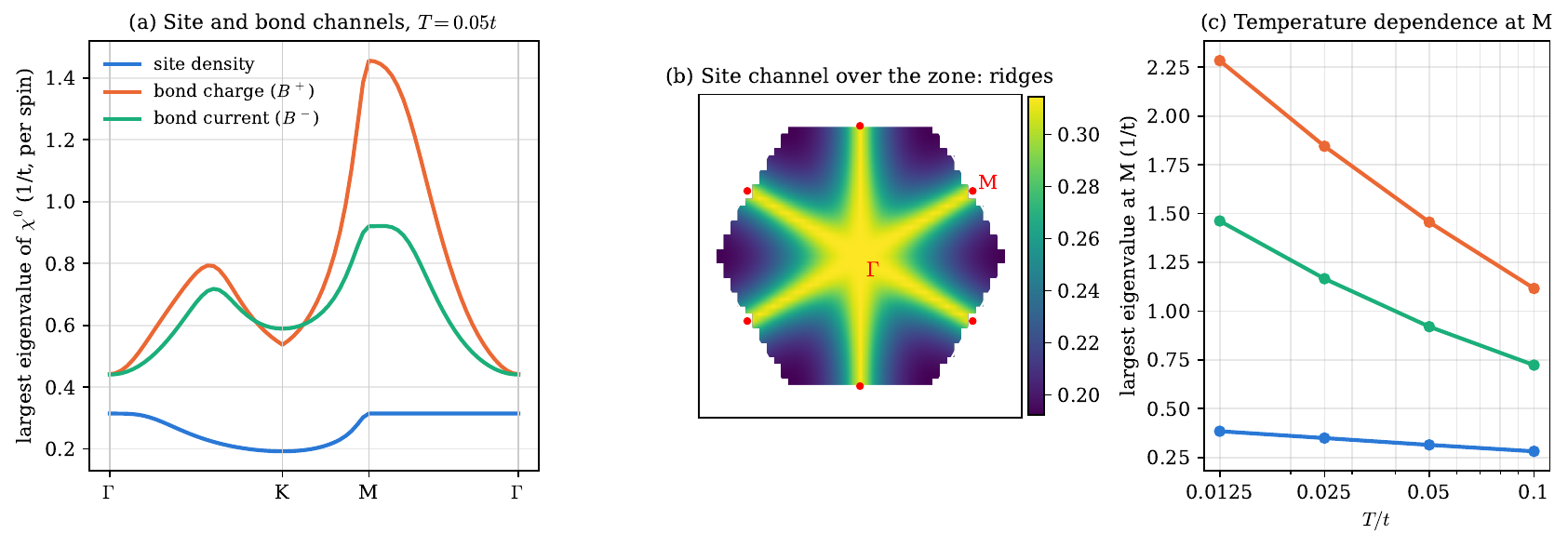}
\caption{Non-interacting susceptibilities at the upper (p-type) van Hove filling. (a) Largest eigenvalue of the site, bond-charge and bond-current blocks along $\Gamma$--K--M--$\Gamma$, $T=0.05t$. (b) Largest site eigenvalue over the Brillouin zone: ridges along the $\Gamma$--M lines. (c) Values at M versus temperature, with the colors of (a): site $\propto\ln(1/T)$, bond $\sim\ln^2(1/T)$.}
\label{fig:chi0}
\end{figure*}

Figure~\ref{fig:chi0} shows the result at the upper van Hove filling, where the saddle points are of p-type (pure) sublattice character. In the site-charge channel the largest eigenvalue is \emph{exactly constant} along the whole $\Gamma$--M line and forms three ridges over the Brillouin zone, as noted in Ref.~\cite{Wang2013}; it grows only as $\ln(1/T)$, with the same value at M and $\Gamma$. Sliding nesting explains why entire lines are enhanced: every $\vq$ on a line maps a face pair onto itself, giving a one-dimensional-like $\ln(1/T)$ enhancement. At M the van Hove saddle points are additionally connected, and the logarithmically divergent density of states at the saddle points would multiply the nesting logarithm, turning $\ln(1/T)$ into $\ln^2(1/T)$; but at this p-type filling the two saddle points connected by M reside on different sublattices, so the site matrix element between them vanishes (sublattice interference). At the lower van Hove filling ($E_F=-2t$), where the saddle points are of m-type (mixed) sublattice character, the connected saddle points share a sublattice and this suppression is absent. The flatness itself is a structural property of the critical mode, the antisymmetric combination of the two sublattices on the chains perpendicular to the line [$F=\mathrm{diag}(0,1,-1)/\sqrt2$ for the line $\vq\cdot\va_3=0$]: its polarizability $\chi^0_{FF}$ is constant along the line at every filling, with or without nesting. What nesting adds at the upper van Hove filling is the $\ln(1/T)$ enhancement, which makes this flat mode the leading site response. It remains the leading mode for $-1.5t\lesssim E_F\lesssim0$. Outside this window, including the lower (m-type) van Hove filling, another mode dominates and the leading site response is no longer flat; at the lower filling it is peaked at M, where sublattice interference is absent. In the bond channel, by contrast, the operators connect different sublattices and escape sublattice interference: the bond-charge polarizability is strongly peaked at M, about 4.6 times the site value at $T=0.05t$, and grows as $\ln^2(1/T)$, the signature of nesting combined with a van Hove singularity.

\subsection{Two forms of the overdamped boson}
\label{sec:forms}

Near a charge instability the effective interaction between electrons is dominated by a soft, Landau-damped collective mode. We describe it by the overdamped boson model,
\begin{equation}
\begin{gathered}
W(\vq,\omega)=-\frac{V(\vq)}{\omega_m(\vq)-i\omega},\\
B(\vq,\omega)\equiv-\frac1\pi\,\mathrm{Im}\,W(\vq,\omega)=\frac{V(\vq)}{\pi}\frac{\omega}{\omega^2+\omega_m^2},
\end{gathered}
\label{eq:boson}
\end{equation}
where $B(\vq,\omega)$ is the boson spectral function, $V(\vq)\ge0$ is concentrated near the soft region, and $\omega_m$ is the characteristic frequency, at which $B$ peaks. The negative sign is the attractive charge-channel interaction obtained near a charge instability. Electrons couple to it through the vertex of Eq.~\eqref{eq:vertex}, with $F$ the form factor of the critical fluctuation (site charge or bond), as given in Appendix~\ref{app:tb}. Since the low-energy physics involves only the band that crosses the Fermi level, we keep its diagonal element, $g\equiv g^F_{nn}$; a calculation that keeps the full $3\times3$ structure, including interband scattering, gives the same spectra to within 0.02 (Sec.~\ref{sec:rpaB}). The pattern is normalized to unit length, so $g$ is dimensionless and carries only the momentum dependence of the coupling. Its magnitude, which for a lattice-driven fluctuation contains the electron--phonon matrix element, is absorbed into $V(\vq)$, and the electrons feel the effective interaction $|g|^2W$. We characterize the strength by the zone-averaged static interaction for unit vertex, $\bar U=\int\frac{d^2q}{(2\pi)^2}|W(\vq,0)|$.

\begin{figure*}[t]
\centering
\includegraphics[width=\textwidth]{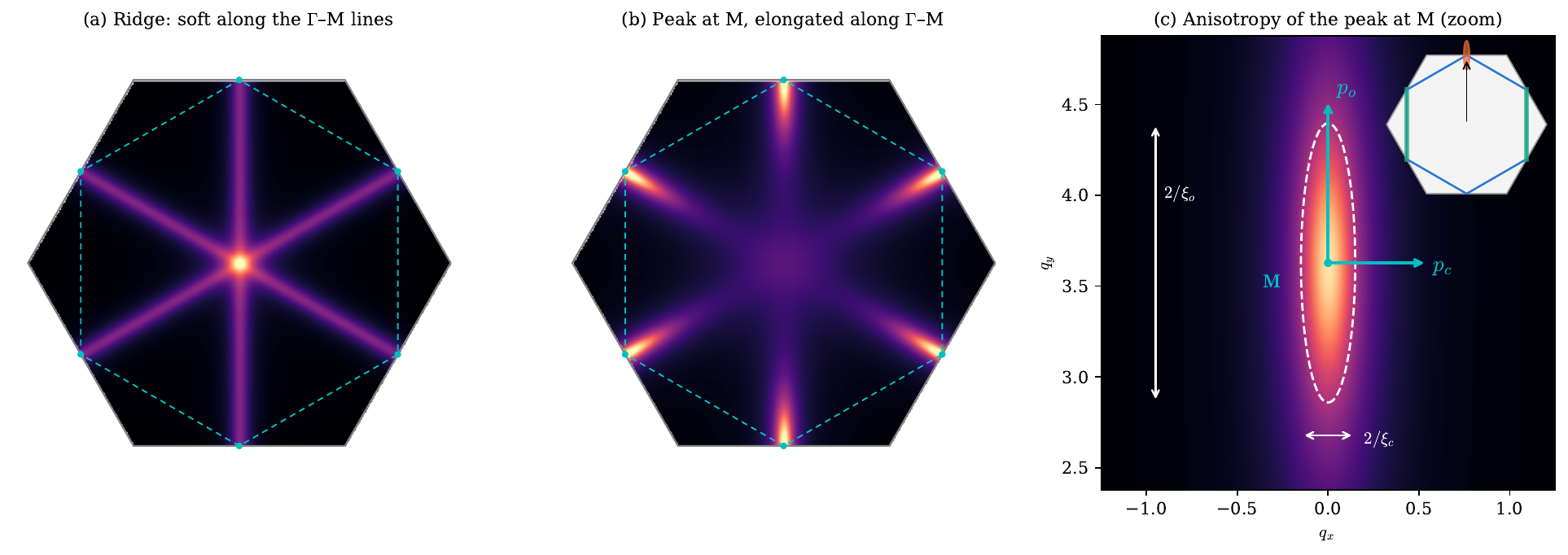}
\caption{The two forms of the overdamped boson (schematic). (a) Ridge: soft along the $\Gamma$--M lines. (b) Peak at M, elongated along $\Gamma$--M. Dashed: Fermi surface. (c) Anisotropy of the peak, Eq.~\eqref{eq:omegap}: zoom on the peak at $\vM=(0,2\pi/\sqrt3)$, with the axes $p_c$ and $p_o$ and the widths $2/\xi_c$ and $2/\xi_o$ (dashed ellipse, $\xi_c/\xi_o\approx5$, as found in Sec.~\ref{sec:rpa}). The inset shows its position relative to the faces $k_x=\pm\pi$ that this $\vM$ nests (green): the broad direction $p_o$ runs parallel to these faces, the narrow direction $p_c$ perpendicular to them.}
\label{fig:forms}
\end{figure*}

Sliding nesting and the channel structure of Fig.~\ref{fig:chi0} lead to two distinct forms of $V(\vq)$ (Fig.~\ref{fig:forms}):
\begin{itemize}
\item \textbf{Ridge} (Fig.~\ref{fig:forms}a). Soft along the entire lines $\vq\cdot\va_i=0$, narrow across them. The correlations are one-dimensional: chains along $\va_i$ are correlated over a length $\xi$ and uncorrelated with each other. This is the site charge instability channel.
\item \textbf{Peak} (Fig.~\ref{fig:forms}b). Centered at the M points but elongated along $\Gamma$--M, with correlation lengths $\xi_c$ along the chains and $\xi_o<\xi_c$ between them (the subscripts c and o denote directions along and orthogonal to the chains). This is the bond charge instability channel.
\end{itemize}
The bond-charge channel is distinct from the bond-current ($B^-$) channel, which is enhanced at M but subleading (Fig.~\ref{fig:chi0}a). In the rest of the paper we refer to the site-charge and bond-charge channels briefly as the \emph{site} and \emph{bond} channels.
Near the soft momentum $\vQ$, with $\vq=\vQ+\mathbf p$, both forms are described by an anisotropic characteristic frequency,
\begin{equation}
\omega_m(\vQ+\mathbf p)=\omega_m\big(1+\xi_c^2p_c^2+\xi_o^2p_o^2\big),
\label{eq:omegap}
\end{equation}
where $p_c$ is the component of $\mathbf p$ along the chain direction associated with $\vQ$ and $p_o$ the component orthogonal to it, along $\Gamma$--$\vQ$; the static weight $V(\vq)/\omega_m(\vq)$ then has widths $1/\xi_c$ and $1/\xi_o$ in the two directions. The ridge is the limit $\xi_o\to0$. For the peak, an important geometric property is that its broad direction, $p_o$, runs parallel to the faces nested by $\vQ$, while its narrow direction, $p_c$, is perpendicular to them (Fig.~\ref{fig:forms}c). A momentum spread along $p_o$ only slides the partner state along the straight Fermi surface and costs no energy.

The model therefore requires the following parameters: the form (ridge or peak) and the form factor that determines $g$; the strength $\bar U$; the correlation lengths $\xi_c$ and $\xi_o$; and the frequency $\omega_m$. They can be obtained from a microscopic calculation (Sec.~\ref{sec:rpa}), from numerical simulations, or from experiment. Two constraints are useful throughout. First, a sum rule determines the mass enhancement averaged over the Fermi surface independently of $\omega_m$ and $\xi$: $\langle m^*/m-1\rangle\simeq\bar U\,P^{(g)}_Q(0)$, where $P^{(g)}_Q(0)=\sum_\vk\delta(\xi_\vk)\delta(\xi_{\vk+\vQ})|g|^2/\sum_\vk\delta(\xi_\vk)$ is the vertex-weighted nesting density~\cite{WanSavrasov}. Second, when $T\gg\omega_m$ the boson acts on the electrons as a static, random charge modulation, and only its static correlations enter.

\subsection{One-loop self-energy for $\delta$-function interactions}
\label{sec:delta}

In the one-loop approximation the self-energy is
\begin{equation}
\begin{split}
\Sigma(\vk,i\omega_n)={}&-T\sum_{n'}\int\!\frac{d^2q}{(2\pi)^2}\,|g(\vk,\vk+\vq)|^2\\
&\times W(\vq,i\omega_n-i\omega_{n'})\,G_0(\vk+\vq,i\omega_{n'}).
\end{split}
\label{eq:oneloop}
\end{equation}
For the two forms of the boson, the peak and the ridge, Eq.~\eqref{eq:oneloop} can be evaluated in closed form at $T=0$ in the limit where $V(\vq)$ is a $\delta$-function: concentrated at the three M points for the peak, and on the three $\Gamma$--M lines for the ridge. Both results are built from the same elementary contribution: the self-energy produced by scattering, through the overdamped boson, into a single final state $\vk+\vq$, which we call the \emph{partner} of $\vk$, with energy $\xi'=\xi_{\vk+\vq}$. For a boson of strength $\bar U$ concentrated at one momentum transfer and unit vertex, this contribution, $S(\omega;\xi')$, reads, with $u=\omega-\xi'$,
\begin{align}
\mathrm{Re}\,S(\omega;\xi')&=\bar U\omega_m\,\frac{\frac{u}{\pi}\ln\frac{|u|}{\omega_m}-\frac{\omega_m}{2}\,\mathrm{sgn}\,\xi'}{u^2+\omega_m^2},\label{eq:ReS}\\
\mathrm{Im}\,S(\omega;\xi')&=-\bar U\omega_m\,\frac{|u|}{u^2+\omega_m^2}\,\theta(\xi'u),\label{eq:ImS}
\end{align}
obtained from the spectral representation of Eq.~\eqref{eq:oneloop} and a Kramers--Kronig transformation~\cite{WanSavrasov}. The peak gives a self-energy that is a sum of a few such terms, one for each M point; the ridge gives an average of $S$ over a continuum of partners along each line. $\mathrm{Im}\,S$ sets in at $\omega=\xi'$ and rises linearly, and $\mathrm{Re}\,S$ has a logarithmic kink there. Far from the partner energy, $|u|\gg\omega_m$, the state hybridizes with its partner,
\begin{equation}
\mathrm{Re}\,S(\omega;\xi')\simeq\frac{\Delta_{\rm eff}^2}{\omega-\xi'},\qquad \Delta_{\rm eff}^2=\frac{\bar U\omega_m}{\pi}\ln\frac{|u|}{\omega_m},
\label{eq:Deff}
\end{equation}
so that $\Delta_{\rm eff}$ is the scale of the level repulsion. $S$ also gives a mass enhancement $(\bar U/\pi\omega_m)F(|\xi'|/\omega_m)$ with $F(x)=[(x^2-1)\ln x-(x^2+1)+\pi x]/(x^2+1)^2>0$, whose integral over partner energies is independent of $\omega_m$.

\begin{figure*}[t]
\centering
\includegraphics[width=\textwidth]{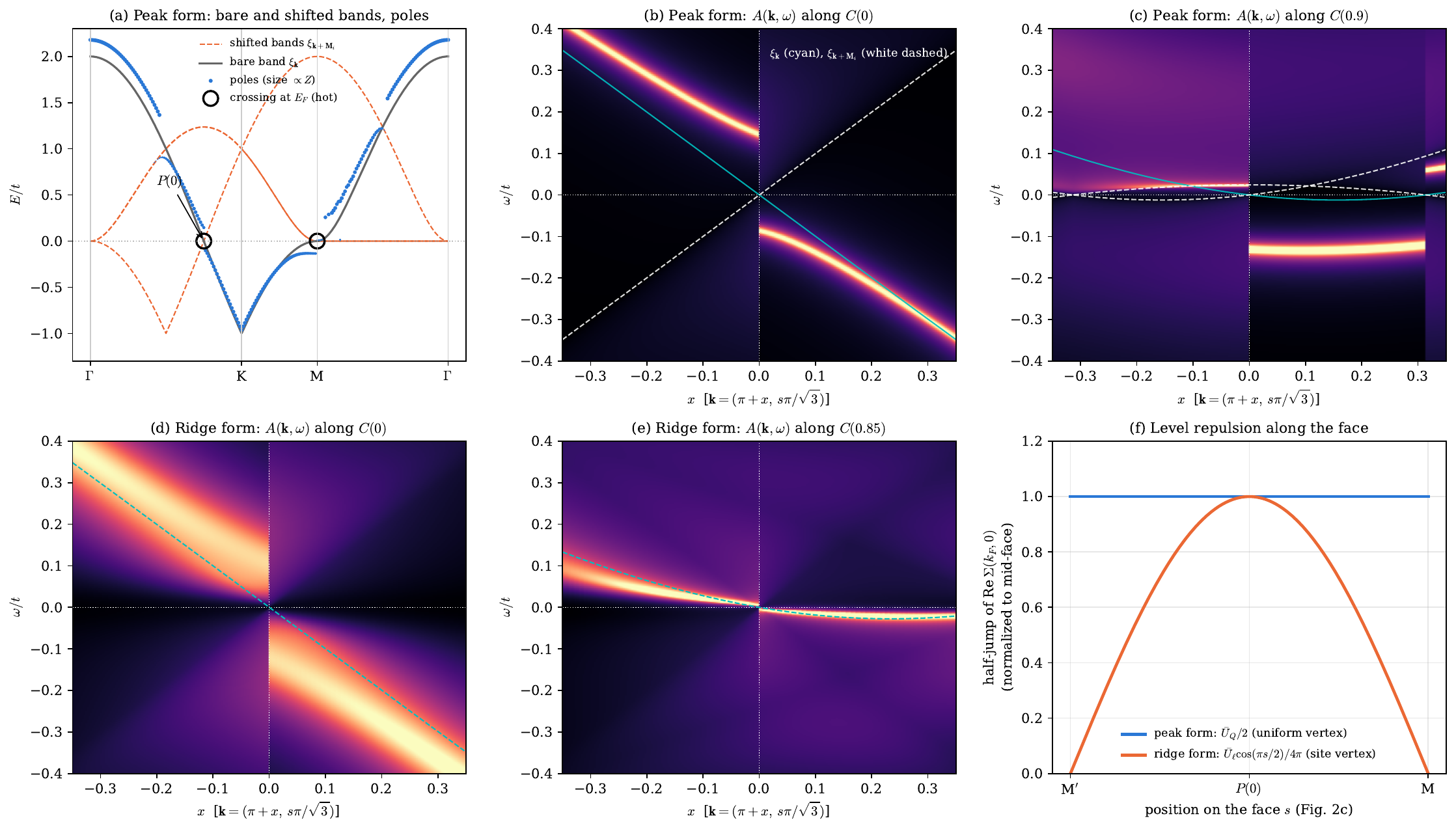}
\caption{One-loop results at $T=0$ for $\delta$-function interactions; the cuts $C(s)$ and the position $s$ along the face are defined in Fig.~\ref{fig:nesting}c. Top row, peak form of instability at M ($\bar U=1.5t$, $\omega_m=0.1t$, uniform vertex): (a) bare band $\xi_\vk$ (gray), the three shifted bands $\xi_{\vk+\vM_i}$ (orange dashed) and the quasiparticle poles (blue, size $\propto Z$) along $\Gamma$--K--M--$\Gamma$; circles mark crossings of the bare band with a shifted copy at the Fermi level, which occur at every point of the Fermi surface; (b,c) spectral function along $C(0)$ and $C(0.9)$ with $\xi_\vk$ (cyan) and $\xi_{\vk+\vM_i}$ (white dashed) overlaid. Bottom row, ridge form of instability ($\bar U_\ell=3t$, $\omega_m=0.08t$, site vertex): (d,e) spectral function along $C(0)$ and $C(0.85)$; here the single shifted band is replaced by the partner continuum of Fig.~\ref{fig:partners}. (f) Half-jump of the static self-energy along the face, normalized to its mid-face value: uniform for the peak form [Eq.~\eqref{eq:jumpM}], $\cos(\pi s/2)$ for the ridge form [Eq.~\eqref{eq:jumpridge}].}
\label{fig:delta}
\end{figure*}

\paragraph{Peak at M.} For $V(\vq)=(2\pi)^2\bar U\omega_m\tfrac13\sum_i\delta(\vq-\vM_i)$ and uniform vertex, $\Sigma(\vk,\omega)=\sum_iS(\omega;\xi_{\vk+\vM_i})$ with $\bar U\to\bar U_Q=\bar U/3$: each $\vk$ has exactly three partners, $\vk+\vM_i$. Because of perfect nesting, every point on a face has one partner exactly at the Fermi level with $\xi_{\vk+\vM}\simeq-\xi_\vk$. The static term then jumps across the Fermi surface,
\begin{equation}
\mathrm{Re}\,\Sigma(\vk,0)\to-\frac{\bar U_Q}{2}\,\mathrm{sgn}\,\xi_{\vk+\vM}=+\frac{\bar U_Q}{2}\,\mathrm{sgn}\,\xi_\vk ,
\label{eq:jumpM}
\end{equation}
which is level repulsion. Figure~\ref{fig:delta}a,b makes this explicit. Because an electron at $\vk$ can scatter only to the three states $\vk+\vM_i$, the renormalized band follows $\xi_\vk$ except near its crossings with the shifted bands $\xi_{\vk+\vM_i}$. Away from the Fermi level these crossings produce kinks or anticrossings. On the Fermi surface the nested copy crosses the bare band exactly at $k_F$, so every Fermi-surface point is a hot spot in the sense of the spin-fermion model~\cite{Hlubina1995}. The quasiparticle pole jumps across the Fermi level, and a (pseudo)gap of order $\bar U_Q/2$ for $\bar U_Q\ll\omega_m$ and $(\bar U_Q\omega_m^2/2)^{1/3}$ for $\bar U_Q\gg\omega_m$ opens along the \emph{whole} Fermi surface (Fig.~\ref{fig:delta}a). At $k_F$ itself $\mathrm{Im}\,\Sigma\simeq-\bar U_Q|\omega|/\omega_m$, so the gap edges are marginally damped. Away from the Fermi level the bands develop kinks where they cross their shifted copies; above a threshold $\bar U_Q\approx52\,\omega_m$, at which the pole equation $E=\xi_\vk+\mathrm{Re}\,\Sigma(\vk,E)$ acquires three solutions at a crossing~\cite{WanSavrasov}, the kinks turn into anticrossings with shadow branches.

\paragraph{Ridge.} For a boson that is a $\delta$-function across each line $\vq\cdot\va_i=0$ and uniform along it, with strength $\bar U_\ell$ per line and the vertex of the critical site mode [for the line $\vq\cdot\va_1=0$, $F=\mathrm{diag}(1,-1,0)/\sqrt2$; Sec.~\ref{sec:rpa}], the kinematics is again exact to first order in the distance $x$ from the face $\vk\cdot\va_1=\pi$. With $\psi=\vq\cdot\va_3$ parametrizing the line and Eq.~\eqref{eq:ustate},
\begin{equation}
\begin{gathered}
\xi_\vk=x\sin\phi,\qquad \xi_{\vk+\vq}=x\sin(\phi+\psi),\\ g=\tfrac{1}{\sqrt2}\cos\!\big(\phi+\tfrac\psi2\big).
\end{gathered}
\end{equation}
For $\vk$ on the face, every partner on the line lies at the Fermi level; off the face, the partner energies sweep $[-|x|,|x|]$. Changing variables to $\beta=\phi+\psi$ and $s'=\sin\beta$ (arcsine-distributed), the self-energy is
\begin{equation}
\Sigma(\vk,\omega)=\frac{\bar U_\ell\omega_m}{4}\int_{-1}^{1}\frac{ds'}{\pi\sqrt{1-s'^2}}\,\big(1-s'\sin\phi\big)\,S(\omega;xs')
\label{eq:ridge}
\end{equation}
(with $S$ per unit strength). In the weight $(1-s'\sin\phi)$, the first term counts partners at $\pm xs'$ equally, that is, partners symmetric about the Fermi level, which damp and renormalize but do not repel; the second term, odd in $s'$, favors partners of opposite energy, which produce the level repulsion, and is proportional to $v_\perp$. Near the Fermi surface,
\begin{equation}
\mathrm{Re}\,\Sigma(k_F^\pm,0)=\pm\frac{\bar U_\ell}{4\pi}\cos\!\Big(\frac{\pi s}{2}\Big),
\label{eq:jumpridge}
\end{equation}
maximal at the face midpoint and \emph{zero at the saddle points}, where only the marginal terms $\mathrm{Re}\,\Sigma\simeq(\bar U_\ell/4\pi\omega_m)\,\omega\ln(|\omega|/\omega_m)$ and $\mathrm{Im}\,\Sigma\simeq-\bar U_\ell|\omega|/8\omega_m$ survive. A full numerical one-loop calculation with the exact band and vertex reproduces Eq.~\eqref{eq:jumpridge} to within 3\%. Far from the Fermi surface the arcsine averages give, at the face midpoint and with $\Delta_{\rm eff}$ of Eq.~\eqref{eq:Deff}, $\Sigma\simeq(\Delta_{\rm eff}^2/4\xi_\vk)\big[1-\sqrt{(z-\xi_\vk)/(z+\xi_\vk)}\big]$: the shadow band of the peak model becomes an inverse-square-root continuum edge, as in one-dimensional fluctuating Peierls systems~\cite{LRA1973,Sadovskii1979}.

\begin{figure*}[t]
\centering
\includegraphics[width=\textwidth]{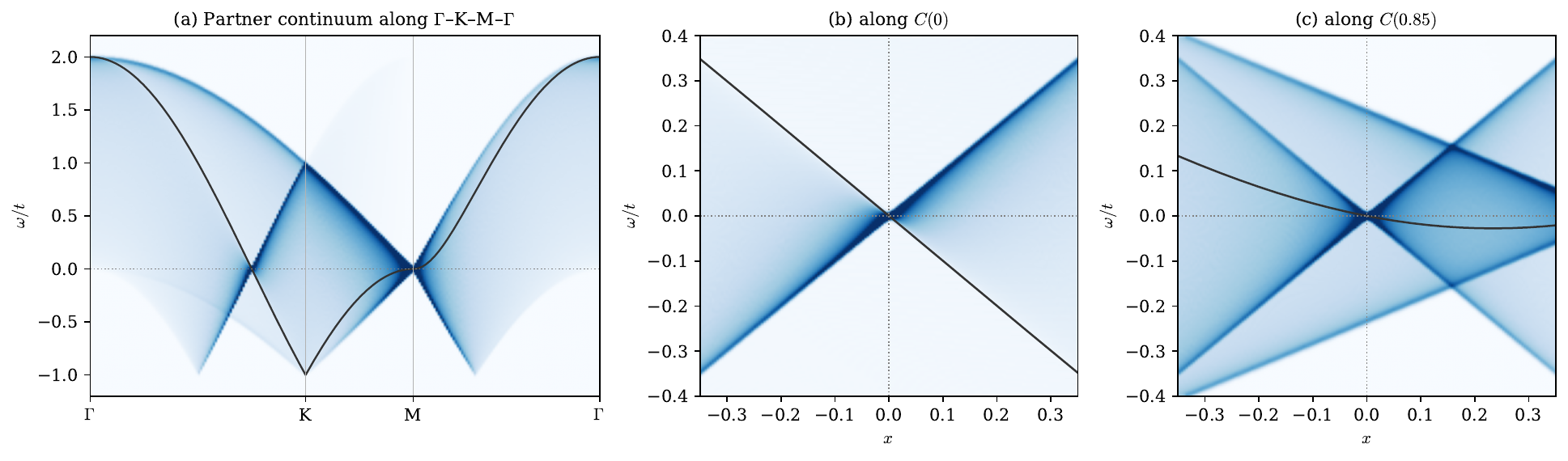}
\caption{Partner continuum for the ridge form of instability, to be contrasted with the single shifted bands of the peak form in Fig.~\ref{fig:delta}a--c: for each $\vk$, the density of partner energies $\xi_{\vk+\vq}$ for $\vq$ on the three $\Gamma$--M lines, weighted by $|g|^2$. (a) Along $\Gamma$--K--M--$\Gamma$; (b) along $C(0)$; (c) along $C(0.85)$, with the cuts defined in Fig.~\ref{fig:nesting}c. The black line is the bare band.}
\label{fig:partners}
\end{figure*}

The single shifted band of the peak model, whose crossing with $\xi_\vk$ defines a hot spot (Fig.~\ref{fig:delta}a,b), is thus replaced by a \emph{partner continuum} (Fig.~\ref{fig:partners}): for each $\vk$ a whole range of partner energies is available, a bow tie pinched at $k_F$ whose weight lies on the edge opposite to the bare band at the face midpoint but is spread over both edges near the corners. The consequences for the single-particle states are opposite in the two models (Fig.~\ref{fig:delta}): the peak gaps the whole Fermi surface including the saddle points, while the ridge with the site vertex produces a pseudogap at the face midpoints and leaves marginal-Fermi-liquid arcs around M.

%=====================================================================
\section{Random-phase approximation}
\label{sec:rpa}
%=====================================================================

\subsection{Site and bond channels: instabilities and boson parameters}
\label{sec:rpaA}

\begin{figure*}[t]
\centering
\includegraphics[width=\textwidth]{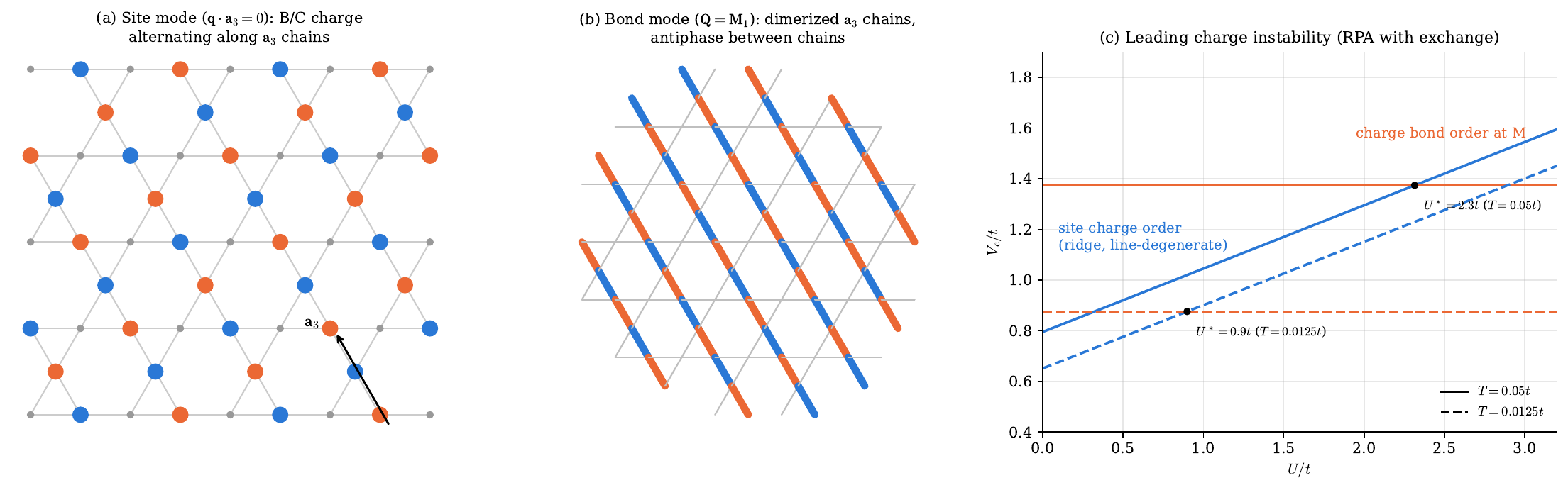}
\caption{RPA with the exchange channel at the upper van Hove filling. (a) Critical site-charge pattern for $\vq$ on the line $\vq\cdot\va_3=0$. (b) Leading bond-order pattern for $\vQ=\vM_1$ at $U=3t$ (orange: strengthened bonds, blue: weakened). (c) Critical $V$ for site charge order (blue) and bond order at M (orange) at two temperatures.}
\label{fig:rpa}
\end{figure*}

We consider the extended Hubbard model $H=H_0+U\sum_in_{i\uparrow}n_{i\downarrow}+V\sum_{\langle ij\rangle}n_in_j$ at the upper van Hove filling. In the site-charge channel the RPA charge susceptibility is
\begin{equation}
\begin{gathered}
\chi_c=2\chi^0\big(1+K\chi^0\big)^{-1},\qquad K(\vq)=U\,\mathbb{I}+2V(\vq),\\
V_{\alpha\beta}(\vq)=2V\cos(\vq\cdot\mathbf d_{\alpha\beta}),
\end{gathered}
\label{eq:rpa}
\end{equation}
where $\chi^0$, $K$ and $V$ are $3\times3$ matrices in sublattice space, $\mathbb I$ is the unit matrix, $\mathbf d_{AB}=\va_1/2$, $\mathbf d_{AC}=\va_2/2$, and $\mathbf d_{BC}=\va_3/2$. On the line $\vq\cdot\va_3=0$ the mirror that exchanges the B and C sublattices leaves $\vq$ invariant, so $\chi^0(\vq)$ and $K(\vq)$ commute with the B$\leftrightarrow$C exchange, and the antisymmetric site pattern $F_S=\mathrm{diag}(0,1,-1)/\sqrt2$ is an exact eigenmode of both. The corresponding eigenvalues, $\chi^0_S\equiv\chi^0_{F_SF_S}$, the largest site-channel eigenvalue at this filling, and $K_{F_SF_S}=U-4V$, are constant along the line, so the instability occurs simultaneously along the whole line at
\begin{equation}
V_c^{\rm site}=\frac{U+1/\chi^0_S}{4}\approx0.80t+\frac U4\qquad(T=0.05t),
\end{equation}
The critical mode alternates B and C charge along each $\va_3$ chain (Fig.~\ref{fig:rpa}a), and RPA cannot select the M point. Our value agrees with the RPA calculation of Ref.~\cite{Wu2024} ($V_c\approx1.1t$ at $U=1.25t$), and renormalization-group studies find that the degeneracy is lifted beyond RPA toward $\vq=0$, an intra-unit-cell (Pomeranchuk) charge order~\cite{Wang2013,Kiesel2013}.

The nearest-neighbor interaction also has an exchange part, $-V\,c^\dagger_ic_jc^\dagger_jc_i=-\tfrac V4[(B^+_{ij})^2+(B^-_{ij})^2]$, attractive in both bond channels. In the $15\times15$ operator space the charge kernel is block diagonal, $U\mathbb I+2V(\vq)$ for site charges and $-V/2$ for bond operators; this is time-dependent Hartree--Fock theory in the charge channel. Because $U$ enters only the site block, the bond instability, $1-\tfrac V2\chi^0_B(\vM)=0$, occurs at $V_c^{\rm bond}\simeq2/\chi^0_B\approx1.38t$ independently of $U$. The critical pattern for $\vQ=\vM_1$ dimerizes the B--C chains along $\va_3$, with antiphase between neighboring chains (Fig.~\ref{fig:rpa}b); superposing the three $\vM_i$ gives star-of-David or tri-hexagonal bond order. The two instabilities cross at
\begin{equation}
U^*=\frac{8}{\chi^0_B}-\frac{1}{\chi^0_S}\approx\begin{cases}2.3t & (T=0.05t),\\ 0.9t & (T=0.0125t),\end{cases}
\end{equation}
above which a bond-charge order at M is the leading charge instability (Fig.~\ref{fig:rpa}c). Because the bond susceptibility grows as $\ln^2(1/T)$, $U^*$ decreases on cooling. The bond order parameter agrees with renormalization-group results~\cite{Wang2013,Kiesel2013} and with the paramagnon-interference mechanism of Ref.~\cite{Tazai2022}, although the driving mechanisms differ: in our treatment the bond order is driven by the exchange part of $V$, without the spin--charge mixing that drives it in those approaches. Two further points qualify this picture. First, the bond channel leads only at low temperature: in the same RPA at $U=3t$ the bond instability precedes the site-charge one below $T\approx t/14$, and at $U=2t$ only below $T\approx t/26$, because a smaller $U$ moves the site-charge instability to smaller $V$. Second, the exchange kernel $-V/2$ acts equally on the spin-antisymmetric bond mode (spin bond order), which mixes with the site block through $-U$ instead of $U+2V(\vq)$. In RPA this mode becomes soft slightly before the charge bond mode at M, by $0.004t$ in $V_c$ at $U=0.5t$ and by up to $0.06t$ at $U=3t$; spin bond order also competes with charge bond order in renormalization-group studies~\cite{Kiesel2013}. We concentrate on the charge bond order, which is the order observed in AV$_3$Sb$_5$; at $U=3t$ the RPA spin channel at M becomes unstable below $T\approx t/29$, which limits the RPA analysis there to higher temperatures.

\begin{figure*}[t]
\centering
\includegraphics[width=\textwidth]{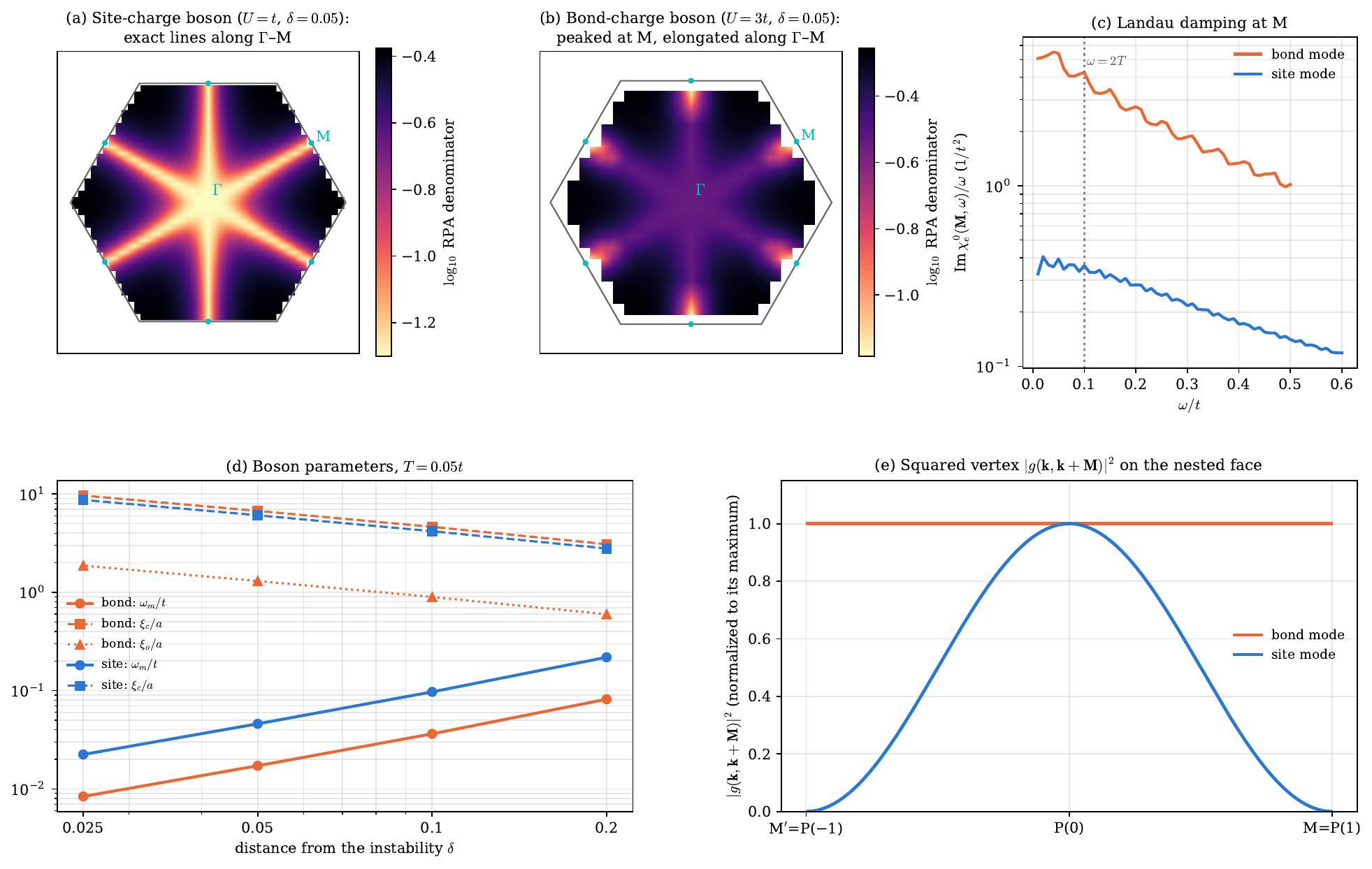}
\caption{The two RPA bosons at $T=0.05t$. (a) RPA denominator of the site-charge channel ($U=t$) and (b) of the bond channel ($U=3t$), both $5\%$ from the instability: ridge and peak forms, as in Fig.~\ref{fig:forms}a,b. (c) Landau damping at M. (d) $\omega_m$ and correlation lengths $\xi_c$, $\xi_o$ [Eq.~\eqref{eq:omegap}] versus the distance $\delta$ from the instability. (e) Squared vertex $|g(\vk,\vk+\vM)|^2$ of Eq.~\eqref{eq:vertex} (band-diagonal element, $g\equiv g^F_{nn}$, with $F$ the critical site or bond pattern) between nested partners $\vk$ and $\vk+\vM$ on the face $k_x=\pi$, with $\vM=(0,2\pi/\sqrt3)$, as a function of the position $s$ (Fig.~\ref{fig:nesting}c), normalized to its maximum.}
\label{fig:bosons}
\end{figure*}

\paragraph{Boson parameters.} Near either instability the response is dominated by the critical pattern $F$, the site pattern $F_S$ of the ridge or the bond pattern $F_B$ of the peak, and both the polarizability and the kernel can be projected onto it: $\chi^0_F\equiv\chi^0_{FF}$ and $K_F\equiv K_{FF}$, with $K_F=U-4V$ in the site channel and $K_F=-V/2$ in the bond channel, both negative near the instability ($\chi^0_S$ and $\chi^0_B$ above are $\chi^0_F$ for the two patterns). The frequency dependence of the critical polarizability is $\chi^0_F(\vq,\omega)\simeq\chi'_F+i\gamma\omega$ for $\omega\lesssim2T$, from Landau damping by the nested faces. Expanding the RPA denominator near $\vQ$ to second order in $\mathbf p$, $1+K_F\chi^0_F\simeq\delta\,(1+\xi_c^2p_c^2+\xi_o^2p_o^2)-i|K_F|\gamma\,\omega$, where $\delta=1+K_F\chi'_F(\vQ)$ is the distance from the instability, the RPA interaction takes exactly the form of Eqs.~\eqref{eq:boson} and \eqref{eq:omegap}, with
\begin{equation}
V(\vq)=\frac{1}{2\gamma},\qquad \omega_m=\frac{\delta}{|K_F|\gamma}.
\label{eq:rpaboson}
\end{equation}
The momentum dependence is carried entirely by $\omega_m(\vQ+\mathbf p)$ of Eq.~\eqref{eq:omegap}; $\omega_m\propto\delta$ and $\xi_{c,o}\propto\delta^{-1/2}$ (dynamical exponent $z=2$). The two bosons have the forms anticipated in Sec.~\ref{sec:forms} (Fig.~\ref{fig:bosons}): the site boson is an exact ridge ($\xi_o=0$), and the bond boson is a peak at M with $\xi_c\approx5\xi_o$. Their parameters are listed in Table~\ref{tab:params}. The bond mode is more strongly damped and softer at equal distance from the instability. Its vertex on the nested faces is \emph{uniform}, corners included, whereas the site vertex vanishes at the saddle points (Fig.~\ref{fig:bosons}e). In both cases the relaxational form of Eq.~\eqref{eq:boson}, which follows from the linear Landau damping $\chi^0_F\simeq\chi'_F+i\gamma\omega$, holds only for $\omega\lesssim2T$. For the zone-averaged strength, $\bar U_i=(V/\omega_m)/(2\pi\xi_c\xi_o)$ per mode for a normalizable (squared-Lorentzian) momentum distribution, with $V$ and $\omega_m$ from Eq.~\eqref{eq:rpaboson}; since the peak height $V/\omega_m\propto1/\delta$ diverges while its area in momentum space, $\propto1/(\xi_c\xi_o)\propto\delta$, shrinks, $\bar U_i$ stays finite as the instability is approached (up to a logarithm for a strictly Ornstein--Zernike form). The boson does not become stronger overall; it becomes taller and narrower, concentrating a fixed total coupling onto the nested states, so that its effects become increasingly momentum-selective while Fermi-surface averages, such as the mass enhancement of the sum rule, remain bounded. This is how the $\delta$-function limits of Sec.~\ref{sec:delta} are reached: at fixed total weight $\bar U_i$, the normalized momentum distribution shrinks as $\delta\to0$, onto the M points for the bond boson ($\xi_c,\xi_o\to\infty$, the peak form) and onto the $\Gamma$--M lines for the site boson ($\xi_c\to\infty$ with $\xi_o=0$, the ridge form). Since $\omega_m\propto\delta$ vanishes at the same time, the instability is approached through the static regime; the $\delta$-function results of Sec.~\ref{sec:delta}, which keep $\omega_m$ finite, describe the case in which the momentum width becomes negligible while $\omega_m$ remains finite. The finite-width, static case is treated in Sec.~\ref{sec:rpaB}.

\begin{table*}[t]
\centering
\caption{Parameters of the two RPA bosons at the upper van Hove filling, $T=0.05t$ (site: $U=t$; bond: $U=3t$). $\delta$ is the distance from the instability.}
\label{tab:params}
\begin{tabular}{llccccc}
\toprule
boson & $\gamma$ ($t^{-2}$) & $\delta$ & $V/t$ & $\omega_m/t$ & $\xi_c/a$ & $\xi_o/a$\\
\midrule
site (ridge), $U=t$ & 0.36 & 0.2 & 0.886 & 0.218 & 2.8 & 0\\
 & & 0.05 & 1.005 & 0.046 & 6.1 & 0\\
 & & 0.0125 & 1.035 & 0.011 & 12.4 & 0\\
\midrule
bond (peak), $U=3t$ & 4.5 & 0.2 & 1.099 & 0.082 & 3.1 & 0.6\\
 & & 0.05 & 1.305 & 0.017 & 6.7 & 1.3\\
 & & 0.025 & 1.339 & 0.008 & 9.6 & 1.9\\
\bottomrule
\end{tabular}
\end{table*}

\subsection{Self-energies and single-particle excitations}
\label{sec:rpaB}

\paragraph{Two regimes.} The one-loop self-energy behaves very differently in the quantum regime ($T\ll\omega_m$: the boson is fast compared with the temperature) and in the classical regime ($T\gg\omega_m$: the boson is slow). In the quantum regime ($T\ll\omega_m$) the static part of $\mathrm{Re}\,\Sigma$ depends on the occupation of the partner state [Eq.~\eqref{eq:ReS}], and only partners of opposite energy repel the level. In the classical regime ($T\gg\omega_m$) the $\nu_l=0$ term of Eq.~\eqref{eq:oneloop} dominates,
\begin{equation}
\begin{gathered}
\Sigma(\vk,\omega)=\Delta_T^2\sum_{i}\Big\langle\frac{|g_i(\vk,\vk+\vQ_i+\mathbf p)|^2}{\omega-\xi_{\vk+\vQ_i+\mathbf p}+i0^+}\Big\rangle_{\mathbf p},\\
\Delta_T^2=T\,\bar U_i,
\end{gathered}
\label{eq:classical}
\end{equation}
the boson acts as a frozen random potential, and \emph{any} partner at the Fermi level, including forward-scattering partners, splits the level into peaks at $\pm\Delta$. Experiments indicate that the charge fluctuations of CsV$_3$Sb$_5$ are in this classical regime (Sec.~\ref{sec:exp}). The same boson can therefore act on different parts of the Fermi surface in the two regimes.

\begin{figure}[t]
\centering
\includegraphics[width=\columnwidth]{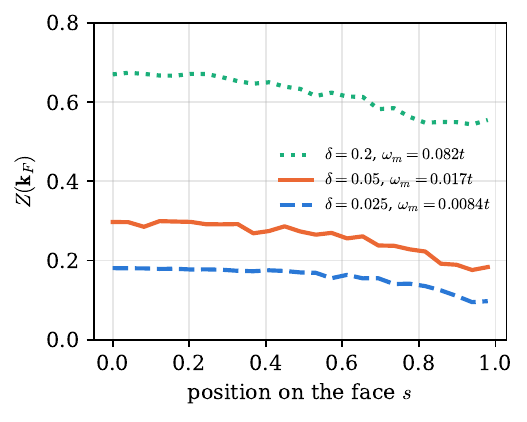}
\caption{Quasiparticle residue along the face ($s$, Fig.~\ref{fig:nesting}c) at $T=0$ (quantum regime) for the RPA bond boson at three distances $\delta$ from the instability, with the parameters of Table~\ref{tab:params} and Eq.~\eqref{eq:rpaboson}, the anisotropic momentum distribution of Eq.~\eqref{eq:omegap} and the bond vertex.}
\label{fig:residue}
\end{figure}

\paragraph{Zero temperature (quantum regime, $T\ll\omega_m$): anisotropic Fermi liquid.} With any finite correlation length the occupation step behind Eqs.~\eqref{eq:jumpM} and \eqref{eq:jumpridge} is smeared over partner energies $\Gamma(\vk)\simeq v_\perp\Delta q$, where $\Delta q\sim1/\xi_c$ is the width of the boson perpendicular to the nested face and $v_\perp$ the partner velocity in that direction. $\mathrm{Re}\,\Sigma(\vk,0)$ becomes continuous and $\mathrm{Im}\,\Sigma(\vk,0)=0$, so a renormalized Fermi surface always survives. The result is a strongly anisotropic Fermi liquid. For the RPA bond boson (Table~\ref{tab:params}, with $V$ and $\omega_m$ from Eq.~\eqref{eq:rpaboson}, the anisotropic momentum distribution of Eq.~\eqref{eq:omegap} and the bond vertex), the quasiparticle residue is smallest at the saddle points and decreases strongly as the instability is approached: from $0.67$ at the face midpoint to $0.56$ at M for $\delta=0.2$, from $0.30$ to $0.18$ for $\delta=0.05$, and from $0.18$ to $0.10$ for $\delta=0.025$ (Fig.~\ref{fig:residue}). The suppression at M reflects the vanishing of $v_\perp$, and with it of $\Gamma$, at the saddle points, while the overall reduction follows the growth of $\bar U/\omega_m$ as the boson softens. These heavy quasiparticles require the boson to be fast compared with the temperature and the energies of interest: for a slow boson the mass renormalization is confined to $|\omega|\lesssim\omega_m$, and at higher energies or temperatures its effect is a scattering rate rather than a mass. The gap of the $\delta$-function models survives only where $\Delta_{\rm eff}\gtrsim v_\perp\Delta q$. Here $\Delta_{\rm eff}$ of Eq.~\eqref{eq:Deff} is the effective hybridization between a state and its partner; it sets the scale of the level repulsion produced by a quantum ($T\ll\omega_m$) boson.

\begin{figure*}[t]
\centering
\includegraphics[width=\textwidth]{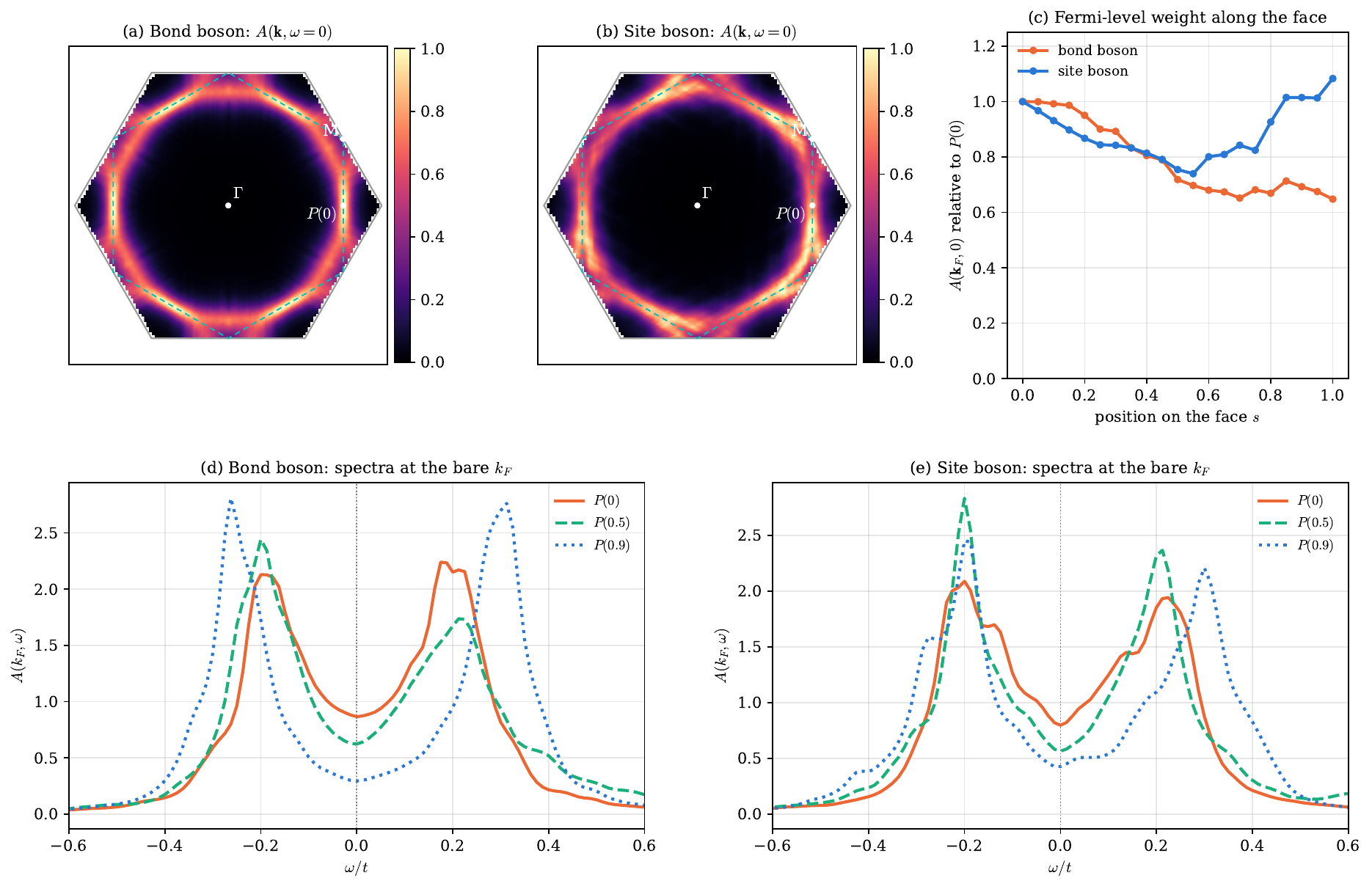}
\caption{Static-regime one-loop spectra for the two RPA bosons ($T=0.05t$, $\Delta_T=0.25t$; bond: $\xi_c=6.7a$, $\xi_o=1.3a$; site: $\xi=6.1a$). Spectral weight at the Fermi level, $A(\vk,0)$, for (a) the bond boson and (b) the site boson; dashed: bare Fermi surface. (c) $A(\vk_F,0)$ along the renormalized Fermi surface, relative to the face midpoint. (d,e) Spectral functions at the bare $k_F$ for $P(0)$, $P(0.5)$ and $P(0.9)$.}
\label{fig:pg}
\end{figure*}

\paragraph{Static regime (classical, $T\gg\omega_m$): pseudogaps and arcs.} In the classical regime the relevant scale is $\Delta_T$ of Eq.~\eqref{eq:classical}, $\Delta_T^2=T\bar U_i$: the root-mean-square amplitude of the static, thermally excited field seen by the electrons. It plays the role of $\Delta_{\rm eff}$ [Eq.~\eqref{eq:Deff}] in the quantum regime: a single partner at the Fermi level splits a level into peaks at $\pm\Delta_T$. These fluctuations produce $\mathrm{Im}\,\Sigma(k_F,0)\approx-\Delta_T^2/\Gamma(\vk)$ and a precursor pseudogap wherever $\Delta_T\gtrsim v_\perp\Delta q$~\cite{Vilk1997,Schmalian1998}, first near the saddle points. Figure~\ref{fig:pg} compares the two RPA bosons in this regime at $T=0.05t$ with the same Fermi-surface-averaged coupling ($\Delta_T=0.25t$) and the chemical potential fixed by the particle number. For the bond boson the energy distribution curves show a two-peak pseudogap along the whole face that deepens and widens toward M: the peaks move from $\pm0.20t$ at $P(0)$ to about $\pm0.29t$ at $P(0.9)$, and the weight at the Fermi level relative to the peaks falls from $0.40$ to $0.11$. On the renormalized Fermi surface the Fermi-level weight decreases monotonically toward M, to about 0.65 of its mid-face value (Fig.~\ref{fig:pg}c). The pseudogap covers the whole face because of the geometry of Fig.~\ref{fig:forms}c, and it is deepest at M because $v_\perp$ vanishes there while the bond vertex does not. For the site boson the bare-$k_F$ spectra also deepen toward M, through forward-scattering partners near $\vq\approx0$ on the ridge, but on the renormalized Fermi surface the weight has a minimum part-way along the face and recovers at M (Fig.~\ref{fig:pg}b,c), and the Fermi surface is deformed by momentum-dependent static shifts. A calculation keeping the full $3\times3$ structure of $\Sigma$ and $G$, including interband scattering, agrees with the band-projected results to within 0.02 in every quantity. At weaker coupling or shorter correlation length the criterion $\Delta_T\gtrsim v_\perp\Delta q$ is met only near the saddle points, and the bond boson leaves Fermi arcs at the face midpoints.

\paragraph{Summary of regimes.} Table~\ref{tab:regimes} collects the results. The location of the strongest effects is controlled by the partner velocity, which vanishes at the saddle points, and by the vertex; for the bond boson both point to M, consistent with the gap structure observed in the charge-ordered state of CsV$_3$Sb$_5$, where the gap is largest at the M saddle points~\cite{Nakayama2021}. Whether a gap or a heavy Fermi liquid results depends on the regime, and the one-loop approximation itself must be tested, since sliding nesting provides each Fermi-surface state with a degenerate manifold of Fermi-level partners. We return to both questions in Sec.~\ref{sec:static}.

\begin{table*}[t]
\centering
\caption{Location and form of the low-energy suppression on the Fermi surface at one loop.}
\label{tab:regimes}
\small
\begin{tabular}{p{5.8cm}lp{6.4cm}}
\toprule
model & regime & result\\
\midrule
peak at M, $\delta$-function, uniform vertex & $T=0$ & gap on the whole Fermi surface\\
ridge, $\delta$-function, site vertex & $T=0$ & pseudogap mid-face; marginal arcs around M\\
finite width & $T=0$ & anisotropic Fermi liquid; heavy quasiparticles near M\\
bond boson (peak), finite width & $T\gg\omega_m$ & pseudogap on the whole face, deepest at M\\
site boson (ridge), finite width & $T\gg\omega_m$ & suppression at bare $k_F$; weight recovers near M\\
\bottomrule
\end{tabular}
\end{table*}

\subsection{Beyond RPA: quantum Monte Carlo}
\label{sec:dqmc}
To test these results beyond RPA we performed determinant quantum Monte Carlo (DQMC) simulations~\cite{BSS1981,White1989,SmoQy2024} of the extended Hubbard model at the upper van Hove filling, on clusters of up to $8\times8$ cells and, at $V=0$, $40\times40$ cells, for $U\le7t$, $V\le t$ and $T\ge t/8$; the sign problem of the non-bipartite lattice limits the accessible region to high temperatures, far from the instability. Earlier DQMC studies of the kagome Hubbard model concentrated on half filling, with its metal--insulator transition and magnetic correlations, and on ferromagnetic tendencies away from it~\cite{MedeirosSilva2022,Correia2026}; the charge fluctuations at van Hove filling had not been examined. The details are given in the Supplemental Material~\cite{SM}. The main results are the following. (i) The site-charge ridge is exactly flat along $\Gamma$--M in all parameter sets, and the bond-charge susceptibility is peaked at M and elongated along $\Gamma$--M, as predicted, though weak and short-ranged in this window. (ii) RPA reproduces the bare bond response and its frequency dependence: at $V=0$ the bond susceptibility at M agrees with RPA to 3\% down to $T=t/6$, and $\chi(\vM,i\nu_1)/\chi(\vM,0)$ to 0.016. (iii) RPA overestimates the nearest-neighbor kernels: the effective coupling is 0.5--0.6 of the RPA exchange kernel in the bond channel and about 0.8 of the RPA kernel in the site-charge channel, with no tendency to approach RPA on cooling, so that within the purely electronic model the bond instability requires a larger $V$ than RPA suggests. (iv) Far from the instability the fluctuations are fast, with rates of several times $T$, as expected from $\omega_m\propto\delta$, and the electrons form an anisotropic Fermi liquid, the zero-temperature row of Table~\ref{tab:regimes}: the Fermi-level weight is reduced at M relative to the face midpoint, the sign predicted by the boson model, by 0.3--3\%, growing on cooling. Matsubara-axis diagnostics (Supplemental Material~\cite{SM}) confirm that a static boson does not describe this window and bound the strength of any static boson there to $\Delta_T\lesssim0.1t$ at $V=0$. DQMC thus confirms the structure of the theory, the ridge and the peak and the sign of the momentum dependence, while the near-critical, quasi-static regime relevant to experiment lies outside its reach.

\section{Static fluctuations in C\lowercase{s}V$_3$S\lowercase{b}$_5$}
\label{sec:static}
%=====================================================================

\subsection{Experimental constraints on the fluctuations}
\label{sec:exp}

The previous sections fixed the \emph{form} of the bosons. Their parameters in a real material, and above all the regime they are in, must come from experiment. Here we collect the evidence for CsV$_3$Sb$_5$, with the parameters of the model in mind.

\paragraph{Energy scale.} A minimal fit of the vanadium kagome bands gives a nearest-neighbor hopping $t\approx0.365$\,eV, with next-nearest-neighbor and interlayer hoppings of about 0.01 and 0.05\,eV~\cite{Barman2024}. The CDW transition at $T_{\rm CDW}\approx94$\,K~\cite{Ortiz2020} thus corresponds to $k_BT_{\rm CDW}\approx8$\,meV$\approx0.022t$. Photoemission finds several van Hove singularities close to the Fermi level, of both p-type (pure) and m-type (mixed) sublattice character~\cite{Kang2022,Hu2022}. Which of them corresponds to the p-type upper filling of the single-orbital model, where sublattice interference suppresses the site-charge channel at M and favors the bond peak, and which to the m-type lower filling, where it does not, has been debated. Recent polarization-dependent photoemission identifies sublattice-pure (p-type) van Hove singularities common to all three compounds, which suppress on-site charge fluctuations and enhance bond fluctuations~\cite{Lan2026}, in line with the bond peak of Sec.~\ref{sec:rpaA}; the ridge form nevertheless remains relevant for m-type singularities.

\paragraph{Ordering vector and dimensionality.} In-plane, the order has the $2\times2$ periodicity of the M points. Along $c$ it is three dimensional: diffuse and inelastic x-ray scattering identify the precursor of the charge order at the L point, that is the M-point in-plane pattern with a $\pi$ phase shift between neighboring layers, and find no precursor at the in-plane M point itself~\cite{Subires2023}. Our two-dimensional model describes the in-plane structure, which is common to both; the interlayer phase, and the coupling between the M- and L-point orders~\cite{Christensen2021}, are additional ingredients that do not change the in-plane geometry of sliding nesting.

\paragraph{The fluctuations are quasi-static.} Three independent observations point to fluctuations that are slow compared with every electronic scale.
\begin{itemize}
\item \emph{A soft, strongly damped lattice mode, or a central peak.} The experimental picture of the lattice dynamics is not settled. Early inelastic x-ray scattering found no phonon softening approaching the transition~\cite{Li2021PRX}, and diffuse and inelastic scattering found the critical scattering dominated by an elastic central peak, with order--disorder character and quasi-static domains that condense through a weakly first-order transition~\cite{Subires2023}. High-resolution inelastic x-ray scattering combined with anharmonic first-principles calculations now reports a soft phonon along M--L, strongest at the L point, where the elastic intensity also grows fastest on cooling, and attributes the transition to this soft-mode instability~\cite{McGuinness2026}; the soft mode near L is reported to become overdamped close to $T_{\rm CDW}$, and its large electron--phonon and anharmonic linewidth would explain why it was not resolved earlier~\cite{Alkorta2023}. For the present purpose the two interpretations lead to the same regime. An overdamped soft mode has the response $\chi(\omega)\propto(\Omega^2-\omega^2-i\omega\Gamma)^{-1}$, which for $\Omega\ll\Gamma$ reduces to the relaxational form of Eq.~\eqref{eq:boson} with $\omega_m=\Omega^2/\Gamma$. As the mode softens, $\omega_m$ vanishes and the fluctuations become quasi-static, whether they are seen as a soft phonon or as a central peak.
\item \emph{Picosecond correlation times.} Ultrafast coherent-phonon spectroscopy on hole-doped CsV$_3$Sb$_{5-x}$Sn$_x$ detects charge-order fluctuations well beyond the doping at which long-range order disappears. Their correlation time is $\tau\approx13$\,ps near the doping-tuned transition and $\approx5$\,ps far from it~\cite{Kongruengkit2025}. This corresponds to $\hbar/\tau\approx0.05$--$0.13$\,meV, between 60 and 160 times smaller than $k_BT_{\rm CDW}$.
\item \emph{Fluctuations persist far from order.} The same fluctuations survive to high doping and high temperature, while long-range order terminates abruptly, suggesting a first-order doping-tuned transition~\cite{Kongruengkit2025}.
\end{itemize}
In the language of Sec.~\ref{sec:forms}, the ultrafast correlation times give $\omega_m\lesssim0.1$\,meV, so $\omega_m/k_BT\sim10^{-2}$ or smaller at all temperatures of interest. The fluctuations are therefore deep in the classical (static) regime of Sec.~\ref{sec:rpaB}. This is two to three orders of magnitude slower than the purely electronic estimate near the instability, $\omega_m\approx0.008$--$0.08t\approx3$--$30$\,meV (Table~\ref{tab:params}). Together with the weak electronic bond kernel found in Sec.~\ref{sec:dqmc}, and with the soft, strongly coupled phonon identified at L~\cite{McGuinness2026}, this points to the lattice as the source of both the slowness and part of the strength. We stress that the experimental interpretation of the lattice dynamics remains under debate~\cite{Li2021PRX,Subires2023,McGuinness2026,Alkorta2023}; what the present analysis requires is only that the fluctuations near the transition be slow compared with the temperature, which both pictures, and the ultrafast data, support.

\paragraph{Correlation lengths and anisotropy.} The precursor scattering is strongly anisotropic: its correlation length is 4--5 unit cells in one direction, nearly constant down to $T_{\rm CDW}$, and less than one unit cell in the other, the scattering forming streaks~\cite{Subires2023}. The streaks include an out-of-plane component, so the comparison with the in-plane ratio $\xi_c/\xi_o\approx5$ of our bond boson is suggestive rather than direct. In Sn-doped samples the in-plane correlation length is already finite at $x=0.025$ ($\xi\approx367$\,\AA), and at $x=0.15$ the correlations become incommensurate and quasi-one-dimensional, along a preferred in-plane axis and correlated over only neighboring vanadium planes ($\xi_L\approx18$\,\AA)~\cite{Kautzsch2023}. Ti substitution on the vanadium sites likewise collapses the three-dimensional order into short-range, quasi-one-dimensional stripe correlations with an anisotropic in-plane correlation length~\cite{Xiao2026,Yang2022}, so that persistent short-range charge correlations are not specific to Sn doping. Quasi-one-dimensional charge correlations are precisely what the chain-like structure of both bosons predicts (Fig.~\ref{fig:rpa}).

\paragraph{Form factor.} A real bond order, the $B^+$ mode of Sec.~\ref{sec:rpaA}, is the common ground of theory~\cite{Wang2013,Kiesel2013,Tazai2022} and is consistent with the gap structure of the ordered state, strongly anisotropic and largest at the M saddle points~\cite{Nakayama2021,Kang2022,Luo2022,Cho2021}. Bond-order fluctuations can in turn mediate a charge-loop-current ($B^-$) order and $Z_3$ nematicity~\cite{Tazai2023}. Whether time-reversal symmetry is also broken, which would add a bond-current ($B^-$) component, is contested: muon-spin relaxation reports it~\cite{Mielke2022,Khasanov2022}, high-resolution Kerr measurements do not~\cite{Saykin2023}.

\paragraph{Strength, and the mass enhancement.} The specific-heat coefficient is about 25\% above its density-functional value~\cite{SpecHeat2024}, and photoemission kinks indicate an electron--phonon coupling of intermediate strength, about twice the density-functional estimate~\cite{Zhong2023,Luo2022}. These numbers measure the mass enhancement, but not the strength of the slow charge fluctuations. The specific heat is measured at low temperature, in the charge-ordered state, where the fluctuations are absent; and even above $T_{\rm CDW}$ a boson with $\omega_m\ll T$ renormalizes the electronic mass only within $|\omega|\lesssim\omega_m$, a sub-meV window here. Above $T_{\rm CDW}$, as below it, the mass enhancement therefore comes mainly from phonons, whose energies of 10--30\,meV are large compared with $\omega_m$. The slow charge fluctuations act instead as quasi-static scatterers: their signature is a momentum-selective scattering rate, largest at the van Hove saddle points, rather than a heavier mass. Their strength is fixed by the Monte Carlo simulations of Sec.~\ref{sec:scmc}, which give a bond-field rms of $0.13$--$0.15t$ above the transition. It can be measured independently by the intensity of diffuse x-ray scattering, which probes the amplitude of the fluctuating distortion, and for the purely electronic part at high temperature DQMC bounds it to $\Delta_T\lesssim0.1t$ (Sec.~\ref{sec:dqmc}).

\begin{table*}[t]
\centering
\caption{Parameters of the static boson model for CsV$_3$Sb$_5$ and their experimental status.}
\label{tab:exp}
\small
\begin{tabular}{p{3.4cm}p{6.6cm}p{5.6cm}}
\toprule
parameter & constraint & status in the model\\
\midrule
$t$ & $0.365$\,eV~\cite{Barman2024}; $k_BT_{\rm CDW}\approx0.022t$ & fixed\\
form and $\vQ$ & in-plane M ($2\times2$), 3D L-point precursor~\cite{Subires2023} & bond peak at M (2D in-plane)\\
form factor & real bond order favored; currents contested~\cite{Mielke2022,Khasanov2022,Saykin2023} & $B^+$; $B^+ + B^-$ as an alternative\\
$\omega_m$ & $\hbar/\tau\approx0.05$--$0.13$\,meV~\cite{Kongruengkit2025}; central peak~\cite{Subires2023} or overdamped soft phonon at L~\cite{McGuinness2026} (debated) & static limit, $\omega_m\ll T$\\
$\xi_c$, $\xi_o$ & 4--5 cells and $<1$ cell (precursor)~\cite{Subires2023}; quasi-1D in doped samples~\cite{Kautzsch2023} & $\xi_c\approx4$--$13a$, $\xi_o\approx1.5$--$7a$ (Sec.~\ref{sec:scmc})\\
strength $\Delta_T$ & not set by the specific heat (mass enhancement mainly from phonons~\cite{SpecHeat2024,Zhong2023}); diffuse x-ray intensity; DQMC bound & calibration: rms $0.13$--$0.15t$ (Sec.~\ref{sec:scmc})\\
\bottomrule
\end{tabular}
\end{table*}

The static boson model for CsV$_3$Sb$_5$ is thus defined up to the strength and the precise correlation lengths, which we treat as scenarios within the ranges of Table~\ref{tab:exp}.

\subsection{Monte Carlo of the effective electron--lattice model}
\label{sec:scmc}

In the static regime the boson acts on the electrons as a frozen, random charge modulation, and the one-loop self-energy, Eq.~\eqref{eq:classical}, is the second-order term in this modulation. This is a strong approximation here: sliding nesting gives every Fermi-surface state a whole manifold of partners at the Fermi level, precisely the situation in which higher orders, the scattering of the partners themselves, matter most. In the static regime, however, the electrons can be treated exactly in each frozen configuration, and the statistics of the configurations can be generated by the electrons themselves. Moreover, near the transition the statistics of the slowly fluctuating bond order need not be Gaussian. This is most transparent in the Ginzburg--Landau description of the transition, which was established for the kagome metals in Refs.~\cite{Denner2021,Park2021,Christensen2021}. These works derived the symmetry-allowed free-energy functional of the three M-point orders and used it to identify the possible $3\vQ$ states, the first-order character of the transition, and the coupling to the out-of-plane order. With $\Phi_i(\mathbf r)$, $i=1,2,3$, the slowly varying amplitudes of the three M-point bond modes, which are real because $e^{i\vM_i\cdot\mathbf R}=\pm1$, the functional reads
\begin{equation}
\begin{aligned}
F[\Phi]=\int d^2r\,\Big[&\sum_i\big(r\,\Phi_i^2+c_c(\partial_c\Phi_i)^2+c_o(\partial_o\Phi_i)^2\big)\\
&+w\,\Phi_1\Phi_2\Phi_3+u\Big(\sum_i\Phi_i^2\Big)^2+v\sum_i\Phi_i^4\Big],
\end{aligned}
\label{eq:GL}
\end{equation}
where $\partial_c$ and $\partial_o$ act along and orthogonal to the chains of mode $i$. Its quadratic part is a Gaussian theory: it fixes the two-point correlations, with $r\propto\delta$ and correlation lengths $\xi_{c,o}^2=c_{c,o}/r$, and it is the static content of the overdamped boson of Secs.~\ref{sec:model} and \ref{sec:rpa}, Eq.~\eqref{eq:omegap}. The cubic term exists only because $\vM_1+\vM_2+\vM_3=\mathbf G$. It favors $3\vQ$ over $1\vQ$ order, makes the transition first order, and selects the sign of $\Phi_1\Phi_2\Phi_3$, that is, the star-of-David or tri-hexagonal pattern; together with the quartic terms, it controls the non-Gaussian character of the fluctuations near the transition. Here, rather than assuming the form of $F$ and its coefficients, we demonstrate with microscopic Monte Carlo simulations what its higher-order structure does to the electrons: we generate the statistics of the bond field by treating it as an effective classical lattice variable in equilibrium with the electrons, so that the electronic free energy supplies all terms of Eq.~\eqref{eq:GL} and beyond.

\paragraph{Free energy of the bond field.} The microscopic counterpart of Eq.~\eqref{eq:GL} is the free energy of effective frozen lattice distortions, with the electrons integrated out. We describe these effective distortions by the bond variables $\phi_b=\alpha u_b$, the relative modulations of the nearest-neighbor hoppings produced by bond displacements $u_b$ through the electron--lattice coupling $\alpha$. For a given configuration $\phi=\{\phi_b\}$ the electrons move in the static Hamiltonian $H[\phi]$, and the free energy of the configuration is
\begin{equation}
F[\phi]=F_{\rm el}[\phi]+\frac\kappa2\sum_b\phi_b^2-f_0\sum_b\phi_b.
\label{eq:scmc}
\end{equation}
The first term is the free energy of the electrons in the configuration $\phi$. The second is the elastic energy of the lattice, with stiffness $K$ entering only through $\kappa=K/\alpha^2$ (in units of $1/t$), the single coupling of the model. The third removes the uniform force exerted by the undistorted electrons, so that $\phi$ measures distortions from equilibrium: the condition $\partial F/\partial\phi_b|_{\phi=0}=0$ gives $f_0=\partial F_{\rm el}/\partial\phi_b|_{\phi=0}$. The electronic free energy, at fixed chemical potential ($E_F=0$), is
\begin{equation}
F_{\rm el}[\phi]=-2T\sum_n\ln\!\big(1+e^{-\beta\varepsilon_n[\phi]}\big),
\label{eq:Fel}
\end{equation}
where $\varepsilon_n[\phi]$ are the eigenvalues of the Hamiltonian
\begin{equation}
H[\phi]=-\sum_b(1-\phi_b)\big(c^\dagger_ic_j+\text{h.c.}\big).
\label{eq:Hphi}
\end{equation}
Configurations are distributed with the Boltzmann weight $P[\phi]\propto e^{-\beta F[\phi]}$. Expanding $F_{\rm el}$ in powers of $\phi$ reproduces the terms of Eq.~\eqref{eq:GL}: the quadratic term is the linear response of the electrons to the distortion, with their polarizability of Sec.~\ref{sec:rpa}, and the cubic and quartic terms are their nonlinear response, so that all coefficients of the functional are generated by the electrons rather than assumed. Treating the lattice classically and the electrons exactly is justified when the fluctuations are slow compared with the temperature, as indicated by experiment (Sec.~\ref{sec:exp}). Monte Carlo simulations of classical fields coupled to electrons that are treated exactly are known as exact-diagonalization Monte Carlo or the static-path approximation; they were developed for the manganites~\cite{Yunoki1998,Dagotto2001}, applied to fluctuating superconducting and charge order~\cite{Mayr2005,Dubi2007}, and extended to large lattices by the travelling-cluster approximation~\cite{Kumar2006}. Since $e^{-\beta F[\phi]}$ is positive for every configuration, it has no sign problem.

\paragraph{Sampling.} We sample $P[\phi]$ by single-bond Metropolis updates. Each update requires the change of $F_{\rm el}$, that is, the eigenvalues $\varepsilon_n$ of $H[\phi]$ before and after the update. Rather than diagonalizing the whole lattice at every step, we obtain them in the travelling-cluster approximation~\cite{Kumar2006}: $H[\phi]$ is diagonalized on a cluster of $(2h+1)\times(2h+1)$ cells centered on the updated bond, which moves with each update, and the change of $F_{\rm el}$ is computed from the cluster eigenvalues. We use $h=4$ in production and $h=3$--6 to check convergence. Replicas at 13--24 temperatures exchange configurations by parallel tempering~\cite{Hukushima1996}, so that the slow dynamics near the first-order-like transition does not trap the simulation. Measurements of the electronic spectra use the eigenvalues and eigenvectors of $H[\phi]$ on the whole lattice, $16\times16$, $24\times24$ and $32\times32$ cells. Independent runs without tempering give the same results.

\begin{figure*}[t]
\centering
\includegraphics[width=\textwidth]{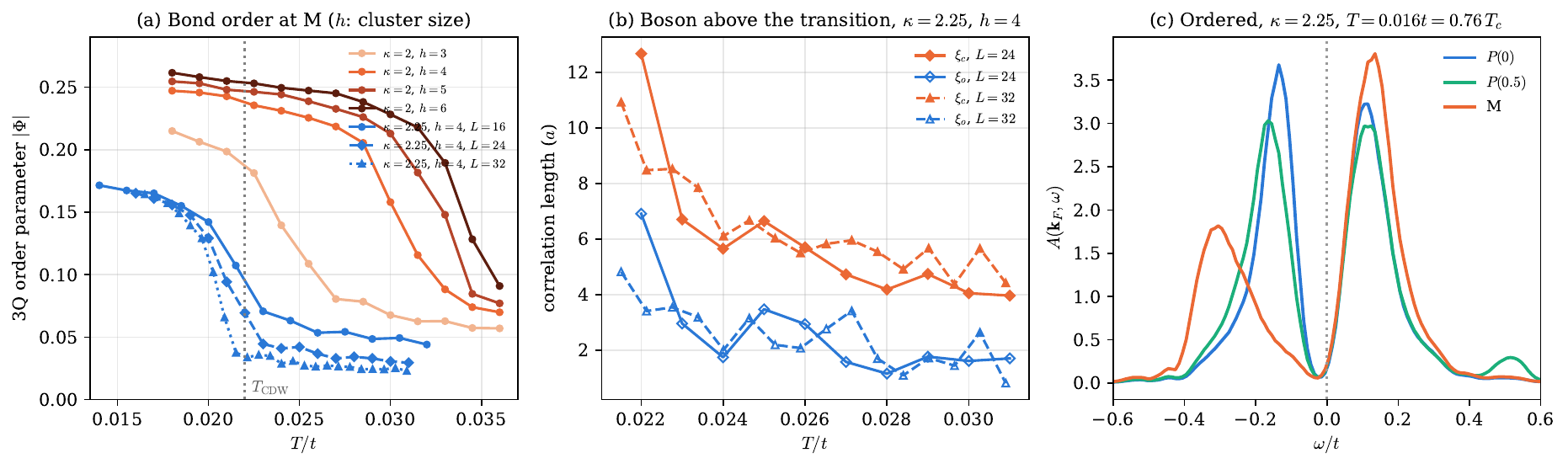}
\caption{Static-field Monte Carlo, Eq.~\eqref{eq:scmc}, with parallel tempering. (a) $3\vQ$ bond-order parameter versus temperature: at $\kappa=2$ on $16\times16$ cells for travelling clusters of size $h=3$--6 (light to dark), showing the convergence in $h$, and at $\kappa=2.25$ with $h=4$ on $16\times16$, $24\times24$ and $32\times32$ cells. Dotted vertical line: $T_{\rm CDW}$ of CsV$_3$Sb$_5$. (b) Correlation lengths of the bond mode above the transition at the calibrated coupling $\kappa=2.25$, $h=4$ (second-moment estimates). (c) Spectral functions on the face in the ordered phase at the calibrated coupling ($\kappa=2.25$, $h=4$, $L=24$, $T=0.016t=0.76\,T_c$).}
\label{fig:scmc}
\end{figure*}

\paragraph{Ordering and calibration.} The electrons drive a $3\vQ$ bond order at the M points (Fig.~\ref{fig:scmc}a); in the ordered phase the lattice-averaged amplitudes $\Phi_i$ of the three M-point modes are nearly equal. In each configuration we project the bond field onto the three M-point bond patterns, which gives $\Phi_i$ and the order parameter $|\Phi|=(\Phi_1^2+\Phi_2^2+\Phi_3^2)^{1/2}$ of Fig.~\ref{fig:scmc}a. We locate the transition at the maximum of its susceptibility, $\chi=(N/T)\big(\langle|\Phi|^2\rangle-\langle|\Phi|\rangle^2\big)$, with $N$ the number of unit cells and $\langle\ldots\rangle$ the Monte Carlo average, which measures the fluctuations of $|\Phi|$. It depends on the size of the travelling cluster but converges: at $\kappa=2$ on $16\times16$ cells, $T_c=0.024t$, $0.030t$, $0.033t$ and $0.0345t$ for $h=3$, 4, 5 and 6, with steps that halve, so that the converged value is about $0.036t$, roughly 20\% above the $h=4$ result. At $\kappa=2.25$ with $h=4$, $T_c=0.0215t$, $0.021t$ and $0.020t$ on $16\times16$, $24\times24$ and $32\times32$ cells, a weak size dependence. The experimental $T_{\rm CDW}\approx0.022t$ (Sec.~\ref{sec:exp}) is thus reproduced by $\kappa=2.25$ at $h=4$, which we use below; correcting for the cluster size, the converged calibration is $\kappa\approx2.3$. Because the electronic results depend on the distance from the transition rather than on the absolute temperature, we present them as functions of $T/T_c$, and we have checked that $\kappa=2$ and $2.25$ give the same behavior on this scale.

\paragraph{The boson above the transition.} Above the ordering temperature the bond structure factor peaks at M (the peak form of Sec.~\ref{sec:forms}), with no ridge along $\Gamma$--M. At the calibrated coupling the field rms is $0.13$--$0.15t$, and on approaching the transition the correlation lengths grow from $\xi_c\approx4a$ and $\xi_o\approx1.5a$ at $1.4\,T_c$ to $\xi_c\approx11$--$13a$ and $\xi_o\approx5$--$7a$ at $1.05\,T_c$, consistently on $24\times24$ and $32\times32$ cells (Fig.~\ref{fig:scmc}b). The anisotropy, $\xi_c/\xi_o\approx2$, is weaker than in the RPA bond boson. Away from the transition these numbers are close to the correlation lengths of 4--5 cells inferred from diffuse x-ray scattering on CsV$_3$Sb$_5$ (Sec.~\ref{sec:exp}). Our Monte Carlo simulations thus provide, from a single calibrated coupling, the boson parameters that the phenomenological model needs.

\begin{figure*}[t]
\centering
\includegraphics[width=\textwidth]{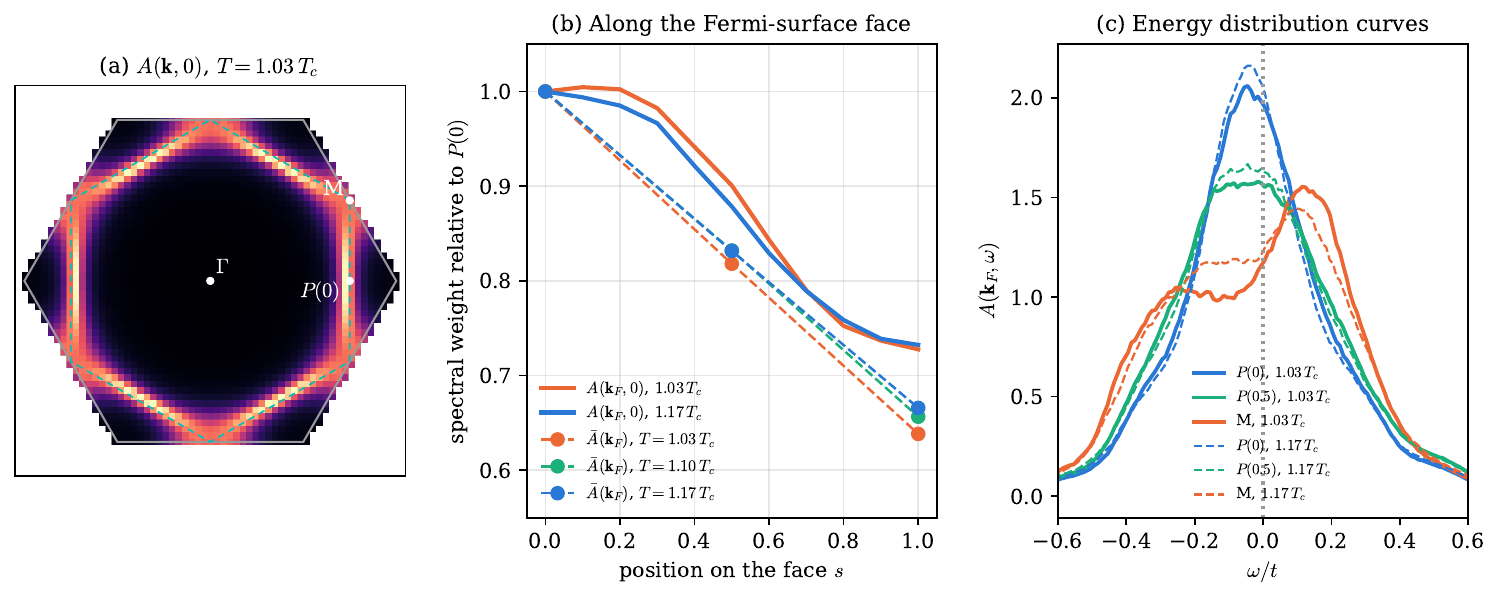}
\caption{Electronic spectra in the fluctuating regime above the bond-order transition, from the static-field Monte Carlo simulations ($\kappa=2$, $h=4$, $L=24$, parallel tempering, $T_c=0.030t$; the calibrated coupling gives the same behavior as a function of $T/T_c$). (a) Spectral weight at the Fermi level, $A(\vk,0)$, at $1.03\,T_c$, averaged over 100 configurations, unfolded to arbitrary momenta with twisted boundary conditions (broadening $0.02t$) and symmetrized over the point group; dashed: bare Fermi surface. (b) Spectral weight along the Fermi-surface face (position $s$, Fig.~\ref{fig:nesting}c), relative to the face midpoint $P(0)$, at $1.03\,T_c$ and $1.17\,T_c$: maximum of $A(\vk,0)$ across the face (lines) and thermally averaged Fermi-level weight $\bar A(\vk_F)$ (symbols). (c) Energy distribution curves at $P(0)$, $P(\tfrac12)$ and M at $1.03\,T_c$ (solid) and $1.17\,T_c$ (dashed).}
\label{fig:scmcel}
\end{figure*}

\paragraph{Electrons.} In the ordered phase the gap covers the whole Fermi surface and is largest at M (Fig.~\ref{fig:scmc}c), consistent with photoemission in the charge-ordered state~\cite{Nakayama2021,Kang2022,Luo2022,Cho2021}. Above the transition the Fermi surface keeps its hexagonal shape, but its spectral weight at the Fermi level is reduced at the M corners on the whole surface (Fig.~\ref{fig:scmcel}a,b), as in the one-loop results of the static regime [Eq.~\eqref{eq:oneloop}, Fig.~\ref{fig:pg}]: with 100 configurations and an energy broadening of $0.02t$, $A(\vk_F,0)$ at M is 0.73--0.75 of its mid-face value between $1.03\,T_c$ and $1.17\,T_c$. To obtain $A(\vk,0)$ at arbitrary momenta we repeat each configuration of the full $L\times L$ lattice periodically, as a supercell, and use twisted boundary conditions, which unfolds the spectral function onto a continuous momentum grid. The thermally averaged Fermi-level weight, $\bar A(\vk_F)=\int d\omega\,A(\vk_F,\omega)(-\partial f/\partial\omega)$, falls from the face midpoint to M to about 0.62--0.70 of its mid-face value (Fig.~\ref{fig:scmcel}b). Away from the transition this momentum selectivity appears as broadening: a single peak on the whole face, broader at M. Close to the transition, however, a dip opens at M while the face midpoint and $P(\tfrac12)$ keep a single peak (Fig.~\ref{fig:scmcel}c): at the calibrated coupling, $A(\vM,0)/A_{\max}=0.61$ at $1.05\,T_c$, $0.81$ at $1.1\,T_c$ and $\approx0.9$ above $1.15\,T_c$, against $\ge0.95$ at the face midpoint. The pseudogap at M is therefore partial and confined to a window of about 10--15\% above the transition.

\paragraph{Random bond fields and the one-loop approximation.} For comparison we also computed the spectra exactly in static random bond fields generated by a Gaussian distribution with spatial correlations peaked at M, with correlation lengths $\xi_c$ and $\xi_o$, using the kernel polynomial method~\cite{Weisse2006}, which obtains $A(\vk,\omega)=\langle\vk|\delta(\omega-H)|\vk\rangle$ from a Chebyshev expansion built by sparse matrix--vector products on the Bloch state $|\vk\rangle$, without diagonalization, on lattices of up to $96\times96$ cells, for $\xi_c=4$--$10a$ and rms modulations of $0.06$--$0.25t$, which include the rms values found in the Monte Carlo simulations above. In all cases the exact spectral function on the Fermi surface keeps a single peak at the Fermi level, $A(\vk_F,0)/A_{\max}\ge0.81$, while the one-loop approximation from the same configurations gives a deep pseudogap; the fluctuations instead broaden the states, about three times more strongly at M than at the face midpoint. Sliding nesting is responsible: to all orders the many Fermi-level partners are themselves scattered into the same degenerate manifold, and the problem resembles a random matrix on that manifold rather than the one-dimensional fluctuating-gap model, in which electrons move in a static random Peierls potential with Gaussian statistics, where each state has a single partner and a pseudogap survives~\cite{LRA1973,Sadovskii1979,Bartosch1999}. Gaussian static fluctuations therefore preserve the momentum selectivity of the bond boson, but as incoherence rather than as a gap. The method, the full parameter scan and the spectra are given in the Supplemental Material~\cite{SM}.

\subsection{Origin of the pseudogap}
\label{sec:criterion}

\begin{figure*}[t]
\centering
\includegraphics[width=\textwidth]{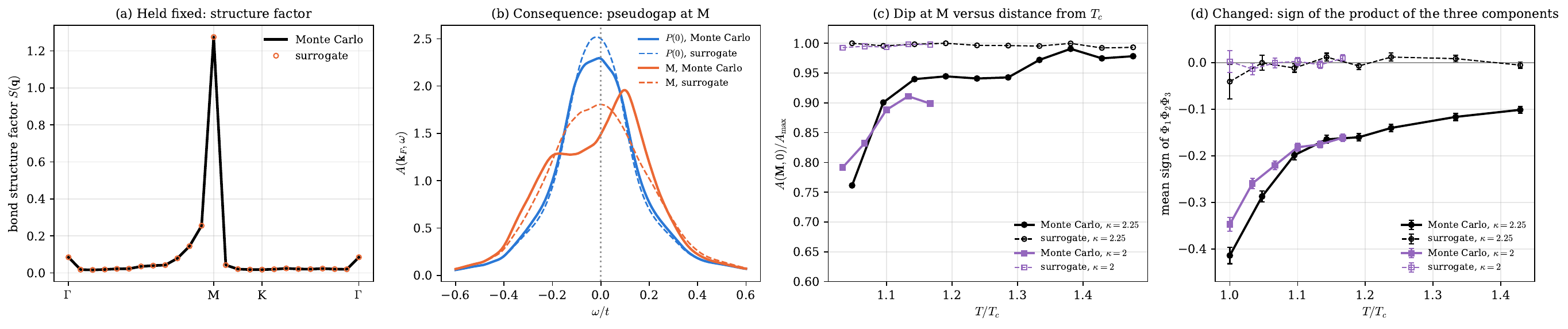}
\caption{Surrogate test ($h=4$, $L=24$, parallel tempering, 101 configurations and three surrogates of each per temperature; panels a and b at the calibrated coupling $\kappa=2.25$, $T_c=0.021t$). (a) Held fixed: the leading eigenvalue of the bond structure factor along $\Gamma$--M--K--$\Gamma$ is identical in the Monte Carlo configurations (line) and in their phase-randomized surrogates (symbols). (b) Consequence: spectral functions at $P(0)$ and M at $1.05\,T_c$, Monte Carlo (solid) and surrogate (dashed). (c) Depth of the dip, $A(\vM,0)/A_{\max}$, versus $T/T_c$ for $\kappa=2.25$ ($T_c=0.021t$) and $\kappa=2$ ($T_c=0.030t$): Monte Carlo (solid) and surrogate (dashed). (d) Changed: mean sign of the product $\Phi_1\Phi_2\Phi_3$ of the three local M-point components, coarse-grained over $3\times3$ cells, versus $T/T_c$, in the Monte Carlo configurations (solid) and in their surrogates (dashed), for $\kappa=2.25$ and $\kappa=2$.}
\label{fig:surrogate}
\end{figure*}

\paragraph{Surrogate test: what is held fixed and what changes.} The Monte Carlo configurations produce a pseudogap at M that static random bond fields generated by a Gaussian distribution, with comparable strength and correlation lengths, do not. To identify which property of the fluctuations is responsible, we compare the spectra computed exactly in stored Monte Carlo configurations with those in \emph{surrogate} configurations obtained from them by randomizing their Fourier phases~\cite{Theiler1992}: every Fourier component of the bond field on all six bonds of the unit cell is multiplied by the same random phase, which preserves the correlations between different bonds and hence the bond pattern of the M-point modes. This holds the structure factor $S_{jj'}(\vq)$ fixed, including the amplitudes of the three M modes; it agrees with that of the original configurations to machine precision (Fig.~\ref{fig:surrogate}a). What changes is the higher-order statistics: in the surrogates the fluctuations become Gaussian, fully determined by their structure factor. At the calibrated coupling and $T=1.05\,T_c$ on $24\times24$ cells, with 101 configurations and three surrogates of each, the dip at M is present in the Monte Carlo configurations ($A(\vM,0)/A_{\max}=0.76$) and absent in the surrogates ($A(\vM,0)/A_{\max}=1$), while the face midpoint is unaffected (1.00 and 0.99; Fig.~\ref{fig:surrogate}b). (These values include a Lorentzian broadening of $0.03t$, which makes the dip shallower than in the run-averaged spectra of Sec.~\ref{sec:scmc}.) Plotted against $T/T_c$, the dip deepens smoothly as the transition is approached, in the same way at $\kappa=2$ and $\kappa=2.25$, while the surrogates remain at 0.98--1.00 throughout (Fig.~\ref{fig:surrogate}c and Supplemental Material~\cite{SM}). The pseudogap at M is therefore not determined by the strength or the correlation lengths of the fluctuations, but by their higher-order statistical structure. The rest of this section explains which structure matters and why.

\paragraph{Local picture.} Near the transition the fluctuations are slow and correlated over a length $\xi$. An electron whose coherence length $v/\Delta$ is shorter than $\xi$ sees an essentially uniform distortion, so the averaged spectrum is a superposition of the spectra of locally ordered patches. A pseudogap at a momentum $\vk$ survives the averaging only if almost every patch gaps the states at $\vk$. The condition $\xi\gg v/\Delta$ is most easily met at the van Hove saddle points, where $v\to0$.

\paragraph{Pairs versus triangles.} What a patch does to the states at $\vk$ depends on the multiplet of states that the order couples together. For one-dimensional Peierls order, or for a single ordering vector connecting hot spots, the states form pairs, with a local Hamiltonian proportional to $\left(\begin{smallmatrix}0&\Phi\\ \Phi^*&0\end{smallmatrix}\right)$ and eigenvalues $\propto\pm|\Phi|$: the gap closes only where the local amplitude $\Phi$ vanishes. The M points of the hexagonal Brillouin zone behave differently, because the ordering vectors close into a loop, $\vM_1+\vM_2+\vM_3=\mathbf G$. The three saddle-point states $|\vM_1\rangle$, $|\vM_2\rangle$, $|\vM_3\rangle$ form a triangle in which each pair is connected by one component of the order: the local amplitude $\Phi_i$ of Eq.~\eqref{eq:GL} couples the two saddle points other than $\vM_i$, with matrix element $g_{\rm M}\Phi_i$, where $g_{\rm M}$ is the vertex of Eq.~\eqref{eq:vertex} between saddle-point states. The local Hamiltonian and its characteristic equation are
\begin{equation}
\begin{gathered}
H_{\triangle}=g_{\rm M}\begin{pmatrix}0&\Phi_3&\Phi_2\\ \Phi_3&0&\Phi_1\\ \Phi_2&\Phi_1&0\end{pmatrix},\\
\lambda^3-g_{\rm M}^2\big(\Phi_1^2+\Phi_2^2+\Phi_3^2\big)\lambda-2g_{\rm M}^3\,\Phi_1\Phi_2\Phi_3=0 .
\end{gathered}
\label{eq:triangle}
\end{equation}
A state remains at the Fermi level ($\lambda=0$) exactly when the product $\Phi_1\Phi_2\Phi_3$ vanishes, that is, whenever \emph{any one} component is locally small. When all three coexist with equal amplitude $\Phi$, the levels are $-\Delta,-\Delta,2\Delta$ or $\Delta,\Delta,-2\Delta$ with $\Delta=g_{\rm M}|\Phi|$, depending on the sign of the product, and the saddle points are fully gapped. The product is the same cubic invariant that appears in the free energy, Eq.~\eqref{eq:GL}. The same three-state structure underlies the mean-field spectrum of the ordered $3\vQ$ state in patch models of the saddle points~\cite{Park2021,Denner2021}; here we use it for fluctuating order. Gaussian fluctuations leave the three components independent, each with a finite probability of being small: in a toy ensemble, the fraction of patches whose local gap is below 5\% of the rms amplitude is 0.3\% for a pair with a complex Gaussian amplitude, but 18\% for the triangle with independent Gaussian components. Fixing only the sign of the product leaves this fraction unchanged, whereas locally coexisting components reduce it to zero.

\paragraph{Evidence in the simulations.} Projecting the stored configurations onto the three M-point bond patterns and coarse-graining over $3\times3$ cells, we find that the Monte Carlo configurations have both more coexistence and a preferred sign of the product than their surrogates, which share all two-point correlations, and that both grow together with the dip at M as the transition is approached. At the calibrated coupling the mean sign of $\Phi_1\Phi_2\Phi_3$ grows from $-0.10\pm0.01$ at $1.4\,T_c$ through $-0.16$ and $-0.20$ to $-0.29\pm0.01$ at $1.05\,T_c$ and $-0.41\pm0.02$ at $T_c$, against $0.00\pm0.02$ in the surrogates at every temperature (Fig.~\ref{fig:surrogate}d), and the weakest of the three local components rises from 0.34 to 0.41 of the rms amplitude, against 0.33--0.37 in the surrogates. The same trend holds at $\kappa=2$, with a mean sign of $-0.35\pm0.01$ at its transition. The modest enhancement of coexistence is consistent with a partial rather than full pseudogap. The ordered state provides a further check: for a negative product, Eq.~\eqref{eq:triangle} predicts an asymmetric splitting of the saddle-point levels, one level at $-2\Delta$ and two at $+\Delta$, and the spectrum at M in Fig.~\ref{fig:scmc}c indeed has a single peak near $-0.30t$ and one near $+0.14t$, consistent with $\Delta\approx0.14t$.

\paragraph{Scope.} The three-state analysis of Eq.~\eqref{eq:triangle} is analytical and relies only on the geometry of the ordering vectors and on the slowness of the fluctuations. Its numerical demonstration, through the surrogate test and the coexistence statistics, is for the kagome model studied here. Its extension to other systems is a prediction: it should apply to any order whose wave vectors close into loops, $\mathbf Q_1+\mathbf Q_2+\mathbf Q_3=\mathbf G$, such as M-point ($2\times2$) charge, bond or spin order on triangular, honeycomb and kagome lattices, including three-component charge order in transition-metal dichalcogenides such as 1$T$-TiSe$_2$~\cite{DiSalvo1976}. For such orders we expect a precursor pseudogap to signal locally coexisting components, that is, fluctuating $3\vQ$ order, rather than the fluctuation strength or correlation length alone, whereas for pair-coupled states Gaussian fluctuations suffice once $\xi\gg v/\Delta$.

%=====================================================================
\section{Conclusion}
\label{sec:conclusions}
%=====================================================================

On the kagome lattice at van Hove filling, sliding nesting and the sublattice structure of the Bloch states give the charge susceptibility near an instability one of two forms, a ridge along $\Gamma$--M or a peak at M. Describing both by an overdamped boson model, we obtained closed-form one-loop self-energies, in which the peak gaps the van Hove saddle points and the ridge the face midpoints. In RPA, the exchange part of the nearest-neighbor interaction produces the peak, a bond-charge instability at M, whose boson suppresses quasiparticles most strongly at the saddle points.

Determinant quantum Monte Carlo confirms the ridge and the peak beyond RPA, although in its accessible high-temperature window, far from the instability, the fluctuations are fast and the electrons remain an anisotropic Fermi liquid. Experiments on CsV$_3$Sb$_5$, whether they are interpreted as a central peak or as an overdamped soft phonon at the L point, indicate instead that the fluctuations near the transition are slow compared with the temperature, which places the material in the static regime.

Treating the static fluctuations exactly, our Monte Carlo simulation of an effective electron--lattice model reproduces the bond order at M and, just above the transition, a partial pseudogap at the saddle points. The pseudogap requires locally ordered domains rather than merely strong fluctuations, for a reason we expect to hold more broadly: when the ordering wave vectors close into a triangle, as for M-point order on triangular, honeycomb and kagome lattices, the local gap is set by the product of the three order-parameter components, so a precursor pseudogap signals their local coexistence. The momentum dependence predicted by the overdamped boson model, with the strongest effects at the van Hove saddle points, survives in all cases; whether it appears as a pseudogap or as incoherence depends on how close the fluctuations are to forming local order.

For CsV$_3$Sb$_5$ these results lead to concrete expectations. Precursor effects of the charge order should be concentrated at the M saddle points in all cases. Far above $T_{\rm CDW}$ they should appear as a momentum-selective broadening and loss of coherence at M rather than as a gap; a pseudogap at M should develop only close to the transition, where locally ordered $3\vQ$ domains form. Momentum-resolved photoemission on doped or strained samples, in which the fluctuations extend over a wide range of temperature, together with diffuse x-ray scattering of the same samples, which measures the correlation lengths and their anisotropy, can test this connection directly.

Finally, the same fluctuations bear on superconductivity. Charge fluctuations concentrated on part of the Fermi surface can mediate pairing when they are fast enough, as found for Ba$_{1-x}$K$_x$BiO$_3$~\cite{WanSavrasov} and for CsV$_3$Sb$_5$ under pressure~\cite{TianSavrasov}. Since their characteristic frequency drops as the charge-ordering transition is approached, the fluctuations are quasi-static near the transition and do not contribute appreciably to pairing at ambient pressure; when pressure suppresses the order, however, they become dynamic and contribute substantially to pairing, which raises the superconducting $T_c$. Both are regimes of the same boson.

\begin{acknowledgments}
The work was supported by US DOE Grant No. DE-SC0026106.
\end{acknowledgments}

\section*{Data Availability}
The data and the Monte Carlo code are available from the authors upon reasonable request; the determinant quantum Monte Carlo code SmoQyDQMC is publicly available~\cite{SmoQy2024}.

\appendix
\section{Tight-binding model, Bloch states and form factors}
\label{app:tb}

\paragraph{Lattice and Bloch Hamiltonian.} The kagome lattice has primitive vectors $\va_1=(1,0)$ and $\va_2=(\tfrac12,\tfrac{\sqrt3}{2})$, with $\va_3=\va_2-\va_1$ (lattice constant $a=1$), and sublattices A, B, C at $\mathbf r_A=0$, $\mathbf r_B=\va_1/2$ and $\mathbf r_C=\va_2/2$. We use the Fourier convention $c_{\vk a}=N^{-1/2}\sum_{\mathbf R}e^{-i\vk\cdot(\mathbf R+\mathbf r_a)}c_{\mathbf Ra}$, in which the phases refer to the actual site positions. With nearest-neighbor hopping $-t$ ($t>0$) the Bloch Hamiltonian is
\begin{equation}
H(\vk)=-2t\begin{pmatrix}0&\cos k_1&\cos k_2\\ \cos k_1&0&\cos k_3\\ \cos k_2&\cos k_3&0\end{pmatrix},\qquad k_n=\tfrac12\vk\cdot\va_n,
\label{eq:Hk}
\end{equation}
which is real and symmetric in this gauge, so the Bloch vectors $u^\mu(\vk)$ can be chosen real. Its eigenvalues are $E_{\rm flat}=2t$ and
\begin{equation}
E_\pm(\vk)=-t\Big[1\mp\sqrt{3+2\textstyle\sum_n\cos(\vk\cdot\va_n)}\,\Big].
\label{eq:bands}
\end{equation}

\paragraph{Fermi surface and Bloch states.} At either van Hove filling the condition $E_\pm=E_{\rm vH}$ reduces to $\sum_n\cos(\vk\cdot\va_n)=-1$, which is satisfied on the straight lines $\vk\cdot\va_i=\pm\pi$. On the face $\vk\cdot\va_1=\pi$, with $\phi=\vk\cdot\va_3$ running along the face [$\phi=-\tfrac\pi2(1-s)$ at $P(s)$], the upper-band state at the Fermi level is
\begin{equation}
u(\vk)=\big(\cos\tfrac\phi2,\ \sin\tfrac\phi2,\ 0\big),
\label{eq:ustate}
\end{equation}
which becomes a pure A or a pure B state at the two corners of the face (p-type); the states on the other faces follow by symmetry.

\paragraph{Form factors.} For the density $n_\alpha$ on sublattice $\alpha$, and for the bond-charge and bond-current operators $B^+_b=c^\dagger_ic_j+\text{h.c.}$ and $B^-_b=i(c^\dagger_ic_j-\text{h.c.})$ on a bond $b$ that joins a site $j$ of sublattice $\beta$ to the site $i=j+\mathbf d_b$ of sublattice $\alpha$, the form factors of $O_F(\vq)=\sum_{\mathbf x}e^{i\vq\cdot\mathbf x}(\dots)$, with $\mathbf x$ the site or bond-midpoint position, are
\begin{equation}
\begin{aligned}
F^{(n_\alpha)}_{ab}&=\delta_{a\alpha}\delta_{b\alpha},\\
F^{(B^+_b)}_{ab}&=\delta_{a\alpha}\delta_{b\beta}\,e^{-i(\vk+\vq/2)\cdot\mathbf d_b}+\delta_{a\beta}\delta_{b\alpha}\,e^{i(\vk+\vq/2)\cdot\mathbf d_b},\\
F^{(B^-_b)}_{ab}&=i\,\delta_{a\alpha}\delta_{b\beta}\,e^{-i(\vk+\vq/2)\cdot\mathbf d_b}-i\,\delta_{a\beta}\delta_{b\alpha}\,e^{i(\vk+\vq/2)\cdot\mathbf d_b}.
\end{aligned}
\label{eq:formfactors}
\end{equation}
There are six bonds per unit cell, $\mathbf d_b=\pm\va_1/2$ (A--B), $\pm\va_2/2$ (A--C) and $\pm\va_3/2$ (B--C), so Eq.~\eqref{eq:formfactors} defines fifteen operators. A general fluctuation pattern is a normalized combination $F=\sum_mh_mF^{(m)}$ with $\sum_m|h_m|^2=1$; a site-charge pattern has the diagonal form factor $F=\mathrm{diag}(e_A,e_B,e_C)$.


\begin{thebibliography}{99}
\bibitem{Guo2009} H.-M. Guo and M. Franz, \textit{Topological insulator on the kagome lattice}, Phys.\ Rev.\ B \textbf{80}, 113102 (2009).
\bibitem{Ko2009} W.-H. Ko, P. A. Lee, and X.-G. Wen, \textit{Doped kagome system as exotic superconductor}, Phys.\ Rev.\ B \textbf{79}, 214502 (2009).
\bibitem{Yu2012} S.-L. Yu and J.-X. Li, \textit{Chiral superconducting phase and chiral spin-density-wave phase in a Hubbard model on the kagome lattice}, Phys.\ Rev.\ B \textbf{85}, 144402 (2012).
\bibitem{Kiesel2012} M. L. Kiesel and R. Thomale, \textit{Sublattice interference in the kagome Hubbard model}, Phys.\ Rev.\ B \textbf{86}, 121105(R) (2012).
\bibitem{Kiesel2013} M. L. Kiesel, C. Platt, and R. Thomale, \textit{Unconventional Fermi surface instabilities in the kagome Hubbard model}, Phys.\ Rev.\ Lett.\ \textbf{110}, 126405 (2013).
\bibitem{Wang2013} W.-S. Wang, Z.-Z. Li, Y.-Y. Xiang, and Q.-H. Wang, \textit{Competing electronic orders on kagome lattices at van Hove filling}, Phys.\ Rev.\ B \textbf{87}, 115135 (2013).
\bibitem{Mazin2014} I. I. Mazin, H. O. Jeschke, F. Lechermann, H. Lee, M. Fink, R. Thomale, and R. Valent\'i, \textit{Theoretical prediction of a strongly correlated Dirac metal}, Nat.\ Commun.\ \textbf{5}, 4261 (2014).
\bibitem{Ortiz2019} B. R. Ortiz, L. C. Gomes, J. R. Morey, M. Winiarski, M. Bordelon, J. S. Mangum, I. W. H. Oswald, J. A. Rodriguez-Rivera, J. R. Neilson, S. D. Wilson, E. Ertekin, T. M. McQueen, and E. S. Toberer, \textit{New kagome prototype materials: discovery of KV$_3$Sb$_5$, RbV$_3$Sb$_5$, and CsV$_3$Sb$_5$}, Phys.\ Rev.\ Materials \textbf{3}, 094407 (2019).
\bibitem{Ortiz2020} B. R. Ortiz, S. M. L. Teicher, Y. Hu, J. L. Zuo, P. M. Sarte, E. C. Schueller, A. M. M. Abeykoon, M. J. Krogstad, S. Rosenkranz, R. Osborn, R. Seshadri, L. Balents, J. He, and S. D. Wilson, \textit{CsV$_3$Sb$_5$: A $\mathbb{Z}_2$ topological kagome metal with a superconducting ground state}, Phys.\ Rev.\ Lett.\ \textbf{125}, 247002 (2020).
\bibitem{Ortiz2021K} B. R. Ortiz, P. M. Sarte, E. M. Kenney, M. J. Graf, S. M. L. Teicher, R. Seshadri, and S. D. Wilson, \textit{Superconductivity in the $\mathbb{Z}_2$ kagome metal KV$_3$Sb$_5$}, Phys.\ Rev.\ Materials \textbf{5}, 034801 (2021).
\bibitem{Yin2021} Q. Yin, Z. Tu, C. Gong, Y. Fu, S. Yan, and H. Lei, \textit{Superconductivity and normal-state properties of kagome metal RbV$_3$Sb$_5$ single crystals}, Chin.\ Phys.\ Lett.\ \textbf{38}, 037403 (2021).
\bibitem{Neupert2022} T. Neupert, M. M. Denner, J.-X. Yin, R. Thomale, and M. Z. Hasan, \textit{Charge order and superconductivity in kagome materials}, Nat.\ Phys.\ \textbf{18}, 137 (2022).
\bibitem{Jiang2022} K. Jiang, T. Wu, J.-X. Yin, Z. Wang, M. Z. Hasan, S. D. Wilson, X. Chen, and J. Hu, \textit{Kagome superconductors AV$_3$Sb$_5$ (A = K, Rb, Cs)}, Natl.\ Sci.\ Rev.\ \textbf{10}, nwac199 (2022).
\bibitem{Wilson2024} S. D. Wilson and B. R. Ortiz, \textit{AV$_3$Sb$_5$ kagome superconductors}, Nat.\ Rev.\ Mater.\ (2024).
\bibitem{Kang2022} M. Kang, S. Fang, J.-K. Kim, B. R. Ortiz, S. H. Ryu, J. Kim, J. Yoo, G. Sangiovanni, D. Di Sante, B.-G. Park, C. Jozwiak, A. Bostwick, E. Rotenberg, E. Kaxiras, S. D. Wilson, J.-H. Park, and R. Comin, \textit{Twofold van Hove singularity and origin of charge order in topological kagome superconductor CsV$_3$Sb$_5$}, Nat.\ Phys.\ \textbf{18}, 301 (2022).
\bibitem{Hu2022} Y. Hu, X. Wu, B. R. Ortiz, S. Ju, X. Han, J. Ma, N. C. Plumb, M. Radovic, R. Thomale, S. D. Wilson, A. P. Schnyder, and M. Shi, \textit{Rich nature of van Hove singularities in kagome superconductor CsV$_3$Sb$_5$}, Nat.\ Commun.\ \textbf{13}, 2220 (2022).
\bibitem{Ortiz2021X} B. R. Ortiz, S. M. L. Teicher, L. Kautzsch, P. M. Sarte, N. Ratcliff, J. Harter, J. P. C. Ruff, R. Seshadri, and S. D. Wilson, \textit{Fermi surface mapping and the nature of charge-density-wave order in the kagome superconductor CsV$_3$Sb$_5$}, Phys.\ Rev.\ X \textbf{11}, 041030 (2021).
\bibitem{Li2021PRX} H. Li, T. T. Zhang, T. Yilmaz, Y. Y. Pai, C. E. Marvinney, A. Said, Q. W. Yin, C. S. Gong, Z. J. Tu, E. Vescovo, C. S. Nelson, R. G. Moore, S. Murakami, H. C. Lei, H. N. Lee, B. J. Lawrie, and H. Miao, \textit{Observation of unconventional charge density wave without acoustic phonon anomaly in kagome superconductors AV$_3$Sb$_5$ (A = Rb, Cs)}, Phys.\ Rev.\ X \textbf{11}, 031050 (2021).
\bibitem{Chen2021} K. Y. Chen, N. N. Wang, Q. W. Yin, Y. H. Gu, K. Jiang, Z. J. Tu, C. S. Gong, Y. Uwatoko, J. P. Sun, H. C. Lei, J. P. Hu, and J.-G. Cheng, \textit{Double superconducting dome and triple enhancement of $T_c$ in the kagome superconductor CsV$_3$Sb$_5$ under high pressure}, Phys.\ Rev.\ Lett.\ \textbf{126}, 247001 (2021).
\bibitem{Yu2021} F. H. Yu, D. H. Ma, W. Z. Zhuo, S. Q. Liu, X. K. Wen, B. Lei, J. J. Ying, and X. H. Chen, \textit{Unusual competition of superconductivity and charge-density-wave state in a compressed topological kagome metal}, Nat.\ Commun.\ \textbf{12}, 3645 (2021).
\bibitem{Zheng2022} L. Zheng, Z. Wu, Y. Yang, L. Nie, M. Shan, K. Sun, D. Song, F. Yu, J. Li, D. Zhao, S. Li, B. Kang, Y. Zhou, K. Liu, Z. Xiang, J. Ying, Z. Wang, T. Wu, and X. Chen, \textit{Emergent charge order in pressurized kagome superconductor CsV$_3$Sb$_5$}, Nature \textbf{611}, 682 (2022).
\bibitem{Mielke2022} C. Mielke III, D. Das, J.-X. Yin, H. Liu, R. Gupta, Y.-X. Jiang, M. Medarde, X. Wu, H. C. Lei, J. Chang, P. Dai, Q. Si, H. Miao, R. Thomale, T. Neupert, Y. Shi, R. Khasanov, M. Z. Hasan, H. Luetkens, and Z. Guguchia, \textit{Time-reversal symmetry-breaking charge order in a kagome superconductor}, Nature \textbf{602}, 245 (2022).
\bibitem{Khasanov2022} R. Khasanov, D. Das, R. Gupta, C. Mielke III, M. Elender, Q. Yin, Z. Tu, C. Gong, H. Lei, E. T. Ritz, R. M. Fernandes, T. Birol, Z. Guguchia, and H. Luetkens, \textit{Time-reversal symmetry broken by charge order in CsV$_3$Sb$_5$}, Phys.\ Rev.\ Research \textbf{4}, 023244 (2022).
\bibitem{Saykin2023} D. R. Saykin, C. Farhang, E. D. Kountz, D. Chen, B. R. Ortiz, C. Shekhar, C. Felser, S. D. Wilson, R. Thomale, J. Xia, and A. Kapitulnik, \textit{High resolution polar Kerr effect studies of CsV$_3$Sb$_5$: Tests for time-reversal symmetry breaking below the charge-order transition}, Phys.\ Rev.\ Lett.\ \textbf{131}, 016901 (2023).
\bibitem{Li2022NP} H. Li, H. Zhao, B. R. Ortiz, T. Park, M. Ye, L. Balents, Z. Wang, S. D. Wilson, and I. Zeljkovic, \textit{Rotation symmetry breaking in the normal state of a kagome superconductor KV$_3$Sb$_5$}, Nat.\ Phys.\ \textbf{18}, 265 (2022).
\bibitem{Nie2022} L. Nie, K. Sun, W. Ma, D. Song, L. Zheng, Z. Liang, P. Wu, F. Yu, J. Li, M. Shan, D. Zhao, S. Li, B. Kang, Z. Wu, Y. Zhou, K. Liu, Z. Xiang, J. Ying, Z. Wang, T. Wu, and X. Chen, \textit{Charge-density-wave-driven electronic nematicity in a kagome superconductor}, Nature \textbf{604}, 59 (2022).
\bibitem{Subires2023} D. Subires, A. Korshunov, A. H. Said, L. S\'anchez, B. R. Ortiz, S. D. Wilson, A. Bosak, and S. Blanco-Canosa, \textit{Order-disorder charge density wave instability in the kagome metal (Cs,Rb)V$_3$Sb$_5$}, Nat.\ Commun.\ \textbf{14}, 1015 (2023).
\bibitem{Denner2021} M. M. Denner, R. Thomale, and T. Neupert, \textit{Analysis of charge order in the kagome metal AV$_3$Sb$_5$ (A = K, Rb, Cs)}, Phys.\ Rev.\ Lett.\ \textbf{127}, 217601 (2021).
\bibitem{Park2021} T. Park, M. Ye, and L. Balents, \textit{Electronic instabilities of kagome metals: Saddle points and Landau theory}, Phys.\ Rev.\ B \textbf{104}, 035142 (2021).
\bibitem{Lin2021} Y.-P. Lin and R. M. Nandkishore, \textit{Complex charge density waves at van Hove singularity on hexagonal lattices: Haldane-model phase diagram and potential realization in the kagome metals AV$_3$Sb$_5$ (A = K, Rb, Cs)}, Phys.\ Rev.\ B \textbf{104}, 045122 (2021).
\bibitem{Feng2021} X. Feng, K. Jiang, Z. Wang, and J. Hu, \textit{Chiral flux phase in the kagome superconductor AV$_3$Sb$_5$}, Sci.\ Bull.\ \textbf{66}, 1384 (2021).
\bibitem{Tan2021} H. Tan, Y. Liu, Z. Wang, and B. Yan, \textit{Charge density waves and electronic properties of superconducting kagome metals}, Phys.\ Rev.\ Lett.\ \textbf{127}, 046401 (2021).
\bibitem{Wu2021} X. Wu, T. Schwemmer, T. M\"uller, A. Consiglio, G. Sangiovanni, D. Di Sante, Y. Iqbal, W. Hanke, A. P. Schnyder, M. M. Denner, M. H. Fischer, T. Neupert, and R. Thomale, \textit{Nature of unconventional pairing in the kagome superconductors AV$_3$Sb$_5$ (A = K, Rb, Cs)}, Phys.\ Rev.\ Lett.\ \textbf{127}, 177001 (2021).
\bibitem{Profe2024} J. B. Profe, L. Klebl, F. Grandi, H. Hohmann, M. D\"urrnagel, T. Schwemmer, R. Thomale, and D. M. Kennes, \textit{Kagome Hubbard model from a functional renormalization group perspective}, Phys.\ Rev.\ Research \textbf{6}, 043078 (2024).
\bibitem{Wu2024} X. Wu, D. Chakraborty, A. P. Schnyder, and A. Greco, \textit{Crossover between electron-electron and electron-phonon mediated pairing on the kagome lattice}, Phys.\ Rev.\ B \textbf{109}, 014517 (2024).
\bibitem{Tazai2022} R. Tazai, Y. Yamakawa, S. Onari, and H. Kontani, \textit{Mechanism of exotic density-wave and beyond-Migdal unconventional superconductivity in kagome metal AV$_3$Sb$_5$ (A = K, Rb, Cs)}, Sci.\ Adv.\ \textbf{8}, eabl4108 (2022).
\bibitem{Kautzsch2023} L. Kautzsch, Y. M. Oey, H. Li, Z. Ren, B. R. Ortiz, G. Pokharel, R. Seshadri, J. Ruff, T. Kongruengkit, J. W. Harter, Z. Wang, I. Zeljkovic, and S. D. Wilson, \textit{Incommensurate charge-stripe correlations in the kagome superconductor CsV$_3$Sb$_{5-x}$Sn$_x$}, npj Quantum Mater.\ \textbf{8}, 37 (2023).
\bibitem{Kongruengkit2025} T. Kongruengkit, A. N. Capa Salinas, G. Pokharel, B. R. Ortiz, S. D. Wilson, and J. W. Harter, \textit{Persistence of charge density wave fluctuations in the absence of long-range order in a hole-doped kagome metal}, arXiv:2508.13290 (2025).
\bibitem{Oey2022} Y. M. Oey, B. R. Ortiz, F. Kaboudvand, J. Frassineti, E. Garcia, R. Cong, S. Sanna, V. F. Mitrovi\'c, R. Seshadri, and S. D. Wilson, \textit{Fermi level tuning and double-dome superconductivity in the kagome metal CsV$_3$Sb$_{5-x}$Sn$_x$}, Phys.\ Rev.\ Materials \textbf{6}, L041801 (2022).
\bibitem{MMP1990} A. J. Millis, H. Monien, and D. Pines, \textit{Phenomenological model of nuclear relaxation in the normal state of YBa$_2$Cu$_3$O$_7$}, Phys.\ Rev.\ B \textbf{42}, 167 (1990).
\bibitem{Monthoux1991} P. Monthoux, A. V. Balatsky, and D. Pines, \textit{Toward a theory of high-temperature superconductivity in the antiferromagnetically correlated cuprate oxides}, Phys.\ Rev.\ Lett.\ \textbf{67}, 3448 (1991).
\bibitem{Monthoux1993} P. Monthoux and D. Pines, \textit{YBa$_2$Cu$_3$O$_7$: A nearly antiferromagnetic Fermi liquid}, Phys.\ Rev.\ B \textbf{47}, 6069 (1993).
\bibitem{Scalapino1986} D. J. Scalapino, E. Loh, Jr., and J. E. Hirsch, \textit{$d$-wave pairing near a spin-density-wave instability}, Phys.\ Rev.\ B \textbf{34}, 8190 (1986).
\bibitem{Scalapino2012} D. J. Scalapino, \textit{A common thread: The pairing interaction for unconventional superconductors}, Rev.\ Mod.\ Phys.\ \textbf{84}, 1383 (2012).
\bibitem{Hlubina1995} R. Hlubina and T. M. Rice, \textit{Resistivity as a function of temperature for models with hot spots on the Fermi surface}, Phys.\ Rev.\ B \textbf{51}, 9253 (1995).
\bibitem{Vilk1997} Y. M. Vilk and A.-M. S. Tremblay, \textit{Non-perturbative many-body approach to the Hubbard model and single-particle pseudogap}, J.\ Phys.\ I France \textbf{7}, 1309 (1997).
\bibitem{Schmalian1998} J. Schmalian, D. Pines, and B. Stojkovi\'c, \textit{Weak pseudogap behavior in the underdoped cuprate superconductors}, Phys.\ Rev.\ Lett.\ \textbf{80}, 3839 (1998).
\bibitem{Abanov2003} Ar. Abanov, A. V. Chubukov, and J. Schmalian, \textit{Quantum-critical theory of the spin-fermion model and its application to cuprates: normal state analysis}, Adv.\ Phys.\ \textbf{52}, 119 (2003).
\bibitem{Abanov2001} Ar. Abanov, A. V. Chubukov, and A. M. Finkel'stein, \textit{Coherent vs. incoherent pairing in 2D systems near magnetic instability}, Europhys.\ Lett.\ \textbf{54}, 488 (2001).
\bibitem{Metlitski2010} M. A. Metlitski and S. Sachdev, \textit{Quantum phase transitions of metals in two spatial dimensions. II. Spin density wave order}, Phys.\ Rev.\ B \textbf{82}, 075128 (2010).
\bibitem{Castellani1995} C. Castellani, C. Di Castro, and M. Grilli, \textit{Singular quasiparticle scattering in the proximity of charge instabilities}, Phys.\ Rev.\ Lett.\ \textbf{75}, 4650 (1995).
\bibitem{Perali1996} A. Perali, C. Castellani, C. Di Castro, and M. Grilli, \textit{$d$-wave superconductivity near charge instabilities}, Phys.\ Rev.\ B \textbf{54}, 16216 (1996).
\bibitem{Caprara2017} S. Caprara, C. Di Castro, G. Seibold, and M. Grilli, \textit{Dynamical charge density waves rule the phase diagram of cuprates}, Phys.\ Rev.\ B \textbf{95}, 224511 (2017).
\bibitem{Arpaia2019} R. Arpaia, S. Caprara, R. Fumagalli, G. De Vecchi, Y. Y. Peng, E. Andersson, D. Betto, G. M. De Luca, N. B. Brookes, F. Lombardi, M. Salluzzo, L. Braicovich, C. Di Castro, M. Grilli, and G. Ghiringhelli, \textit{Dynamical charge density fluctuations pervading the phase diagram of a Cu-based high-$T_c$ superconductor}, Science \textbf{365}, 906 (2019).
\bibitem{Hertz1976} J. A. Hertz, \textit{Quantum critical phenomena}, Phys.\ Rev.\ B \textbf{14}, 1165 (1976).
\bibitem{Millis1993} A. J. Millis, \textit{Effect of a nonzero temperature on quantum critical points in itinerant fermion systems}, Phys.\ Rev.\ B \textbf{48}, 7183 (1993).
\bibitem{LRA1973} P. A. Lee, T. M. Rice, and P. W. Anderson, \textit{Fluctuation effects at a Peierls transition}, Phys.\ Rev.\ Lett.\ \textbf{31}, 462 (1973).
\bibitem{Sadovskii1979} M. V. Sadovskii, \textit{Exact solution for the density of electronic states in a model of a disordered system}, Sov.\ Phys.\ JETP \textbf{50}, 989 (1979).
\bibitem{WanSavrasov} X. Wan and S. Y. Savrasov, \textit{Charge-fluctuation hot lines and superconductivity in Ba$_{1-x}$K$_x$BiO$_3$}, arXiv:2611.28879 (2026).
\bibitem{BSS1981} R. Blankenbecler, D. J. Scalapino, and R. L. Sugar, \textit{Monte Carlo calculations of coupled boson-fermion systems. I}, Phys.\ Rev.\ D \textbf{24}, 2278 (1981).
\bibitem{White1989} S. R. White, D. J. Scalapino, R. L. Sugar, E. Y. Loh, J. E. Gubernatis, and R. T. Scalettar, \textit{Numerical study of the two-dimensional Hubbard model}, Phys.\ Rev.\ B \textbf{40}, 506 (1989).
\bibitem{SmoQy2024} B. Cohen-Stead, S. Malkaruge Costa, J. Neuhaus, A. Tanjaroon Ly, Y. Zhang, R. Scalettar, K. Barros, and S. Johnston, \textit{SmoQyDQMC.jl: A flexible implementation of determinant quantum Monte Carlo for Hubbard and electron-phonon interactions}, SciPost Phys.\ Codebases \textbf{29} (2024).
\bibitem{Zhong2023} Y. Zhong, S. Li, H. Liu, Y. Dong, K. Aido, Y. Arai, H. Li, W. Zhang, Y. Shi, Z. Wang, S. Shin, H. N. Lee, H. Miao, T. Kondo, and K. Okazaki, \textit{Testing electron--phonon coupling for the superconductivity in kagome metal CsV$_3$Sb$_5$}, Nat.\ Commun.\ \textbf{14}, 1945 (2023).
\bibitem{SpecHeat2024} M. J. Grant, Y. Liu, G.-H. Cao, J. A. Wilcox, Y. Guo, X. Xu, and A. Carrington, \textit{Superconducting energy gap structure of CsV$_3$Sb$_5$ from magnetic penetration depth measurements}, arXiv:2411.05611 (2024).
\bibitem{McGuinness2026} P. H. McGuinness, F. Henssler, M. Alkorta, M. J. Graf von Westarp, A. Korshunov, A. Bosak, D. Ishikawa, A. Q. R. Baron, M. Merz, A.-A. Haghighirad, M. G. Vergniory, S.-M. Souliou, R. Heid, I. Errea, and M. Le Tacon, \textit{Soft mode origin of charge ordering in superconducting kagome CsV$_3$Sb$_5$}, Nat.\ Commun.\ \textbf{17}, 4817 (2026).
\bibitem{Alkorta2023} M. Gutierrez-Amigo, D. Dangi\'c, C. Guo, C. Felser, P. J. W. Moll, M. G. Vergniory, and I. Errea, \textit{Phonon collapse and anharmonic melting of the 3D charge-density wave in kagome metals}, arXiv:2311.14112 (2023).
\bibitem{Weisse2006} A. Wei{\ss}e, G. Wellein, A. Alvermann, and H. Fehske, \textit{The kernel polynomial method}, Rev.\ Mod.\ Phys.\ \textbf{78}, 275 (2006).
\bibitem{Bartosch1999} L. Bartosch and P. Kopietz, \textit{Exact numerical calculation of the density of states of the fluctuating gap model}, Phys.\ Rev.\ B \textbf{60}, 15488 (1999).
\bibitem{Yunoki1998} S. Yunoki, J. Hu, A. L. Malvezzi, A. Moreo, N. Furukawa, and E. Dagotto, \textit{Phase separation in electronic models for manganites}, Phys.\ Rev.\ Lett.\ \textbf{80}, 845 (1998).
\bibitem{Dagotto2001} E. Dagotto, T. Hotta, and A. Moreo, \textit{Colossal magnetoresistant materials: the key role of phase separation}, Phys.\ Rep.\ \textbf{344}, 1 (2001).
\bibitem{Mayr2005} M. Mayr, G. Alvarez, C. {\c S}en, and E. Dagotto, \textit{Phase fluctuations in strongly coupled $d$-wave superconductors}, Phys.\ Rev.\ Lett.\ \textbf{94}, 217001 (2005).
\bibitem{Dubi2007} Y. Dubi, Y. Meir, and Y. Avishai, \textit{Nature of the superconductor--insulator transition in disordered superconductors}, Nature \textbf{449}, 876 (2007).
\bibitem{Kumar2006} S. Kumar and P. Majumdar, \textit{A travelling cluster approximation for lattice fermions strongly coupled to classical degrees of freedom}, Eur.\ Phys.\ J.\ B \textbf{50}, 571 (2006).
\bibitem{Theiler1992} J. Theiler, S. Eubank, A. Longtin, B. Galdrikian, and J. D. Farmer, \textit{Testing for nonlinearity in time series: the method of surrogate data}, Physica D \textbf{58}, 77 (1992).
\bibitem{SM} See Supplemental Material at [URL] for the determinant quantum Monte Carlo simulations, for exact calculations in static random bond fields generated by a Gaussian distribution and for the spectra of the surrogate test, which includes Refs.~\cite{BSS1981,White1989,SmoQy2024,Weisse2006,Bartosch1999}.
\bibitem{TianSavrasov} Y. Tian and S. Y. Savrasov, \textit{Calculated unconventional superconductivity via charge fluctuations in kagome metal CsV$_3$Sb$_5$}, Phys.\ Rev.\ B \textbf{111}, 134507 (2025).
\bibitem{Lan2026} Y. Lan, Y. Lei, C. Le, B. R. Ortiz, N. C. Plumb, M. Radovic, X. Wu, M. Shi, S. D. Wilson, and Y. Hu, \textit{Common sublattice-pure van Hove singularities in the kagome superconductors AV$_3$Sb$_5$ (A = K, Rb, Cs)}, Phys.\ Rev.\ Lett.\ \textbf{136}, 016401 (2026).
\bibitem{Luo2022} H. Luo, Q. Gao, H. Liu, Y. Gu, D. Wu, C. Yi, J. Jia, S. Wu, X. Luo, Y. Xu, L. Zhao, Q. Wang, H. Mao, G. Liu, Z. Zhu, Y. Shi, K. Jiang, J. Hu, Z. Xu, and X. J. Zhou, \textit{Electronic nature of charge density wave and electron-phonon coupling in kagome superconductor KV$_3$Sb$_5$}, Nat.\ Commun.\ \textbf{13}, 273 (2022).
\bibitem{Cho2021} S. Cho, H. Ma, W. Xia, Y. Yang, Z. Liu, Z. Huang, Z. Jiang, X. Lu, J. Liu, Z. Liu, J. Li, J. Wang, Y. Liu, J. Jia, Y. Guo, J. Liu, and D. Shen, \textit{Emergence of new van Hove singularities in the charge density wave state of a topological kagome metal RbV$_3$Sb$_5$}, Phys.\ Rev.\ Lett.\ \textbf{127}, 236401 (2021).
\bibitem{Tazai2023} R. Tazai, Y. Yamakawa, and H. Kontani, \textit{Charge-loop current order and $Z_3$ nematicity mediated by bond order fluctuations in kagome metals}, Nat.\ Commun.\ \textbf{14}, 7845 (2023).
\bibitem{Xiao2026} Q. Xiao, X. Liu, Z. Huang, X. Zheng, S. Zhang, H. Chen, H.-J. Gao, Y. Guo, and Y. Peng, \textit{Evolution from three-dimensional charge density wave to one-dimensional stripe order in CsV$_{3-x}$Ti$_x$Sb$_5$}, Phys.\ Rev.\ Research \textbf{8}, L032020 (2026).
\bibitem{Yang2022} H. Yang, Z. Huang, Y. Zhang, Z. Zhao, J. Shi, H. Luo, L. Zhao, G. Qian, H. Tan, B. Hu, K. Zhu, Z. Lu, H. Zhang, J. Sun, J. Cheng, C. Shen, X. Lin, B. Yan, X. Zhou, Z. Wang, S. J. Pennycook, H. Chen, X. Dong, W. Zhou, and H.-J. Gao, \textit{Titanium doped kagome superconductor CsV$_{3-x}$Ti$_x$Sb$_5$ and two distinct phases}, Sci.\ Bull.\ \textbf{67}, 2176 (2022).
\bibitem{MedeirosSilva2022} A. R. Medeiros-Silva, N. C. Costa, and T. Paiva, \textit{Thermodynamic, magnetic, and transport properties of the repulsive Hubbard model on the kagome lattice}, arXiv:2209.01048.
\bibitem{Correia2026} F. Correia, K. Corbett, and E. Khatami, \textit{Finite-temperature ferromagnetic correlations of the kagome lattice Hubbard model}, arXiv:2604.26827 (2026).
\bibitem{Christensen2021} M. H. Christensen, T. Birol, B. M. Andersen, and R. M. Fernandes, \textit{Theory of the charge density wave in AV$_3$Sb$_5$ kagome metals}, Phys.\ Rev.\ B \textbf{104}, 214513 (2021).
\bibitem{DiSalvo1976} F. J. Di Salvo, D. E. Moncton, and J. V. Waszczak, \textit{Electronic properties and superlattice formation in the semimetal TiSe$_2$}, Phys.\ Rev.\ B \textbf{14}, 4321 (1976).
\bibitem{Hukushima1996} K. Hukushima and K. Nemoto, \textit{Exchange Monte Carlo method and application to spin glass simulations}, J.\ Phys.\ Soc.\ Jpn.\ \textbf{65}, 1604 (1996).
\bibitem{Barman2024} C. K. Barman, S.-W. Kim, and Y. Kim, \textit{Stacking-dependent van Hove singularity shifts in three-dimensional charge density waves of kagome metals AV$_3$Sb$_5$ (A = K, Rb, Cs)}, arXiv:2408.10402 (2024).
\bibitem{Nakayama2021} K. Nakayama, Y. Li, T. Kato, M. Liu, Z. Wang, T. Takahashi, Y. Yao, and T. Sato, \textit{Multiple energy scales and anisotropic energy gap in the charge-density-wave phase of the kagome superconductor CsV$_3$Sb$_5$}, Phys.\ Rev.\ B \textbf{104}, L161112 (2021).
\end{thebibliography}
\end{document}